\documentclass[prx,showkeys,showpacs,footnoteinbib,twocolumn,
unsortedaddress]{revtex4-2}

\usepackage{amsmath}
\usepackage{amssymb}
\usepackage{amsthm}
\usepackage{amsfonts}
\usepackage{epic,eepic}
\usepackage[binary-units]{siunitx}
\usepackage[utf8]{inputenc}
\usepackage{stackengine}
\usepackage{graphicx}
\usepackage{hyperref}
\usepackage[capitalise]{cleveref}
\usepackage[]{todonotes}
\usepackage{bbm}
\usepackage{tikz}
\usepackage{circuitikz}
\usepackage{braket}
\usepackage{physics}
\usepackage{fontawesome}
\usepackage{siunitx}
\usetikzlibrary{%
        shapes.geometric,calc,intersections,%
        decorations.markings,decorations.text,decorations,%
        patterns,decorations.pathmorphing,fpu,%
                decorations.pathreplacing,%
        positioning,automata,shapes.multipart,circuits,%
        circuits.ee, circuits.ee.IEC, shapes.gates.ee,%
                arrows.meta,arrows,math,matrix,
		fadings
}

\newcommand{\mytodo}[1]%
{{\todo[inline,backgroundcolor=blue!10!white]{#1}
}}
\newcommand{\me}{\mathrm{e}}
\newcommand{\mi}{\mathrm{i}}
\newcommand{\md}{\mathrm{d}}

\theoremstyle{definition}

\newtheorem*{remark}{Remark}
\newtheorem*{example}{Example}

\usepackage{todonotes}

\begin{document}

\title{Testing quantum Darwinism dependence on observers' resources}

\author{A. Feller$^3$}
\author{B. Roussel$^2$}
\author{A. Pontlevy$^1$}
\author{P. Degiovanni$^1$}

\affiliation{(1) Univ Lyon, Ens de Lyon, Universit\'e Claude Bernard 
Lyon 1, CNRS, Laboratoire de Physique, F-69342 Lyon, France}
\affiliation{(2) Department of Applied Physics, Aalto University, Aalto, Finland}
\affiliation{(3) Univ. Lille, CNRS, Centrale Lille, UMR 9189 CRIStAL, F-59000 Lille, France}

\begin{abstract}
The emergence of an objective classical picture
is the core question of quantum Darwinism. How does
this reconstructed classical picture depends on the resources
available to observers?
In this Letter, we develop an experimentally relevant model of a qubit 
coupled dispersively to a transmission line and use time-frequency
signal processing techniques to understand if and how the emergent classical
picture is changed when we have the freedom to choose the
fragment decomposition and the type of radiation sent to probe
the system. We show the crucial role of correlations in the reconstruction 
procedure and point to the importance of studying the type of 
measurements that must be done to access an objective classical data.

\end{abstract}


\maketitle

%
Our view of the quantum-to-classical transition, originally based on
decoherence~\cite{Zurek-2003-1} has been expanded with the idea of quantum
Darwinism~\cite{Zurek-2009-1}. This new paradigm shifts the focus from 
the system $S$ to the observers that are probing it. The environment 
$E$ is then viewed as a noisy communication channel from which 
the information about the system is broadcasted to a network of 
observers. Despite being inherently many-body, the approach has led 
to many theoretical results~\cite{Zurek-2001,Riedel-2010,Riedel-2011-1,
Riedel-2012,Zurek-2013,Brandao-2015,Knott-2018-1,Zwolak-2014}
as well as experimental tests~\cite{Burke-2010-1,Zurek-2019}. Although
the existence of an objective observable has been shown to follow from
the first principles of quantum theory \cite{Brandao-2015,Knott-2018-1},
understanding if and how a classical image can be reconstructed by the
observers is still an open question
\cite{Ollivier-2022-1,Garcia-Pinto-2021-1}.

The central tenet of quantum Darwinism is to relate the emergence of a
consensus about the classical state of a quantum system to the many-body 
entanglement structure between the observers and the system. 
Zurek~\cite{Ollivier-2004-1,Zurek-2009-1} used the quantum mutual
information as an estimation of the accessible information \cite{Book-Wilde} in
asymptotic quantum Shannon theory \cite{Touil-2022-1} and
identified the existence of a plateau between the system and 
fragments of the environment as a signature of the redundant 
broadcast of information on pointer states, an intuition confirmed 
in~\cite{Le-2019-1,Feller-2021}.

However, the ability to reconstruct a classical image does not only
depends on the interaction between the system and the environment.
First, it also depends on the initial state of the environment, before
its interaction with the system, as demonstrated by recent studies on 
the role of thermalisation~\cite{Le-2021-1} and many-body 
localization~\cite{Mirkin-2021-1}. Second, it also depends on
the choice of a reference frame, that is which degrees of freedom are
accessible to which observer. The two reflect the resources available 
to observers to probe a quantum system on which we could add 
their computational and communication abilities. The foundational question of
the emergence of classicality can then be precised by analyzing to what 
extent it depends on these resources.

\begin{figure}
\centering
\includegraphics{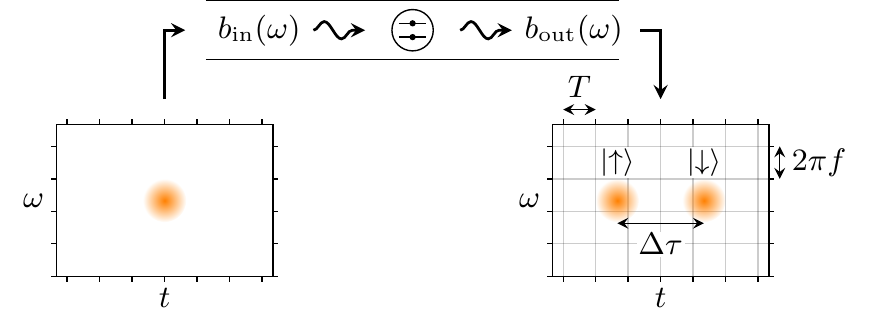}
\caption{\label{fig/the-idea} 
A qubit $S$ is dispersively coupled to a transmission  
line $E$ such that incoming bosonic excitations at given energy 
$\hbar\omega$ are scattered by the qubit in the quantum state 
$\ket{s}$ via the scattering phase $\me^{\mi\theta_s(\omega)}$. 
Fragments are defined from a time-frequency atomic decomposition 
from which information about $\ket{s}$ is recovered.}
\end{figure}

In this Letter, we study this resource dependence of quantum Darwinism. 
We consider a concrete model of a qubit dispersively coupled to a 
transmission line (see \cref{fig/the-idea}).
By using different quantum states for the probe radiation and different
time-frequency windows, 
both the influence of the reference frame and of the probe can be studied. 
We unravel the crucial role played by correlations in the classical
reconstruction process, mostly in a mesoscopic regime where the probe 
involves 1 to 10 photons on average. 


\paragraph*{The dispersive model} The system $S$ is a qubit coupled
dispersively to a transmission line, 
acting as its environment $E$. 
The transmission line is modeled by a 
chiral bosonic field with one chiral mode per pulsation $\omega>0$.
In the dispersive regime, coupling to the qubit is
described by an input/output relation 
$b_{\text{out}}(\omega)=\me^{\mi\theta_s(\omega)}\,b_{\text{in}}(\omega)$ 
where $\theta_s(\omega)$ is a pure phase~\footnote{As expected from
energy conservation in such a 1D chiral system.}.
This is realized in the low-power dispersive regime of circuit-QED 
\cite{Blais-2021-1} used for photon number 
\cite{Gambetta-2006-1,Schuster-2007-1} and qubit state measurements
\cite{Gambetta-2008-1}. 
For simplicity, we consider a time delay model with 
$\theta_s(\omega)=\omega\tau_s$ so that
\begin{equation}
b_{\text{out}}(\omega)= \me^{\mi \omega\tau_s} 
	\,b_{\text{in}}(\omega)
\,.
\label{eq/scattering}
\end{equation}
The two different states of the qubit $\ket{s}$ ($s=0,\ 1$) thus define a time delay 
difference $\Delta\tau=\tau_1-\tau_0$.

\emph{Fragments from atoms of signals}. Studying the emergence
of quantum Darwinism requires defining fragments of the environment 
$E$. We achieve this \emph{via} time-frequency atomic decompositions.
Concretely, an orthonormal basis $\ket{\phi_\alpha}$ of 
$L^2(\mathbb{R})$, $\alpha$ labeling a partition of the time-frequency plane,
defines a quantum reference frame. It corresponds to a 
mode decomposition for the quantum field in which each mode is an atomic 
fragment. Observers are defined 
by aggregating several atomic fragments. 
This enables us to change 
the atomic decomposition, \textit{i.e.} to change the reference frame.

For simplicity, we use the normalized Shannon atoms of signals defined 
as 
$w_{k,l}(\omega)=  \mathbf{1}_{\left[ \frac{2\pi}{T}k,
	\frac{2\pi}{T} (k+1) \right[}(\omega) \, \sqrt{\frac{T}{2\pi}}
	 \,\me^{\mi \omega lT}
\,,$
(see \cref{sec/notations} for normalizations)
with $(l,k)\in\mathbb{Z}\times \mathbb{N}$ providing a regular tessellation of 
the time-frequency half-plane in cells with a bandwidth 
$2\pi/T$ around $\omega_k=2\pi(k+1/2) / T$ and localized in time around
$t_l=lT$. The corresponding
annihilation and creation operators 
$b^{(\dagger)}_{k,l}$, are obtained from the $b^{(\dagger)}_\omega$
operators by a linear
unitary transformation. Since we are dealing with a bosonic
environment, the Fock space factorizes in a tensor product
$
\mathcal{H}_E = \bigotimes_{\omega\geq 0} \mathcal{H}_\omega
	=\bigotimes_{(k,l)\in\mathbb{N}\times\mathbb{Z}} \mathcal{H}_{k,l}
$
of the Fock spaces $\mathcal{H}_{k,l}$ associated with each atomic mode.

The ability to reconstruct a classical image of the system $S$ by
many independent observers in repeated experiments
is quantified by the accessible information
$I_{\text{acc}}(S,F)$ between $S$ and a fragment $F$
\cite{Touil-2022-1}. Apart from few cases \cite{Yao-2014-1}, the
accessible information involves a difficult optimization problem. 
Therefore, we will use the quantum mutual information 
$I(S,F) = S(S) + S(F) - S(SF)$ with $S(X)$ the quantum 
entropy of $X$ as a surrogate estimator \cite{Ollivier-2004-1}. 


\emph{Results} We considered classical and quantum probes.
The classical probe is a multi-mode coherent state
$
\ket{\phi} = \bigotimes_{\omega\geq 0} \ket{q\phi(\omega)}
	=\bigotimes_{(k,l)\in\mathbb{N}\times\mathbb{Z}} \ket{q\phi_{k,l}}
$
with wave-function $\phi_{k,l} = \int_0^{+\infty} 
w_{k,l}^*(\omega) \phi(\omega) \, \md \omega$ and
average photon number $\overline{n}=q^2$. This corresponds to the
illumination of an object with coherent radiation, a pure-state version
of Refs. \cite{Riedel-2010-1,Riedel-2011-1}. Because factorization
of a multi-mode coherent state holds in any reference frame, the
scattered radiation by a given state $\ket{s}$ of $S$ shows
no correlations between the various fragments.
The quantum probe is a Fock state $\ket{n[\phi(\omega)]}
= \frac{\left(a^\dagger[\phi] \right)^n}{\sqrt{n!}}\ket{0}$
with $n$ photons in the $\phi(\omega)$ mode. In contrast to 
coherent states, such a state is highly entangled in terms
of the atomic modes. 
In both cases, the dispersive interaction \cref{eq/scattering} 
amounts to replace the incoming wavefunction by 
$\phi^{(s)}(\omega)=\me^{\mi\omega\tau_s}\phi(\omega)$
conditioned on the state $s$ of the system. 
The signal $\phi(\omega)$ will be conveniently represented
both in the time-domain 
$\phi(t) = \int_{\mathbb{R}} \phi(\omega) 
	\,\me^{\mi \omega t}
	\, \frac{\md \omega}{\sqrt{2\pi}}
$
but also in the time-frequency domain via its Wigner
representation 
$W_\phi(t,\omega) = \int_{\mathbb{R}} \phi^*(t-\tau/2)
\phi(t+\tau/2) \, \me^{\mi\omega\tau} \, \md\tau$.
For definiteness in numerical applications, 
a normalized Gaussian wavepacket
$\phi(\omega) = \frac{1}{\sqrt{\mathcal{N}}} 
	\,\me^{-\frac{(\omega-\omega_0)^2}{2\sigma^2}}
\theta(\omega)$
of width $\sigma$ around a frequency $\omega_0$ is prepared.
Most of the analytic results will not depend on this specific
choice. 

\begin{figure*}
	\includegraphics{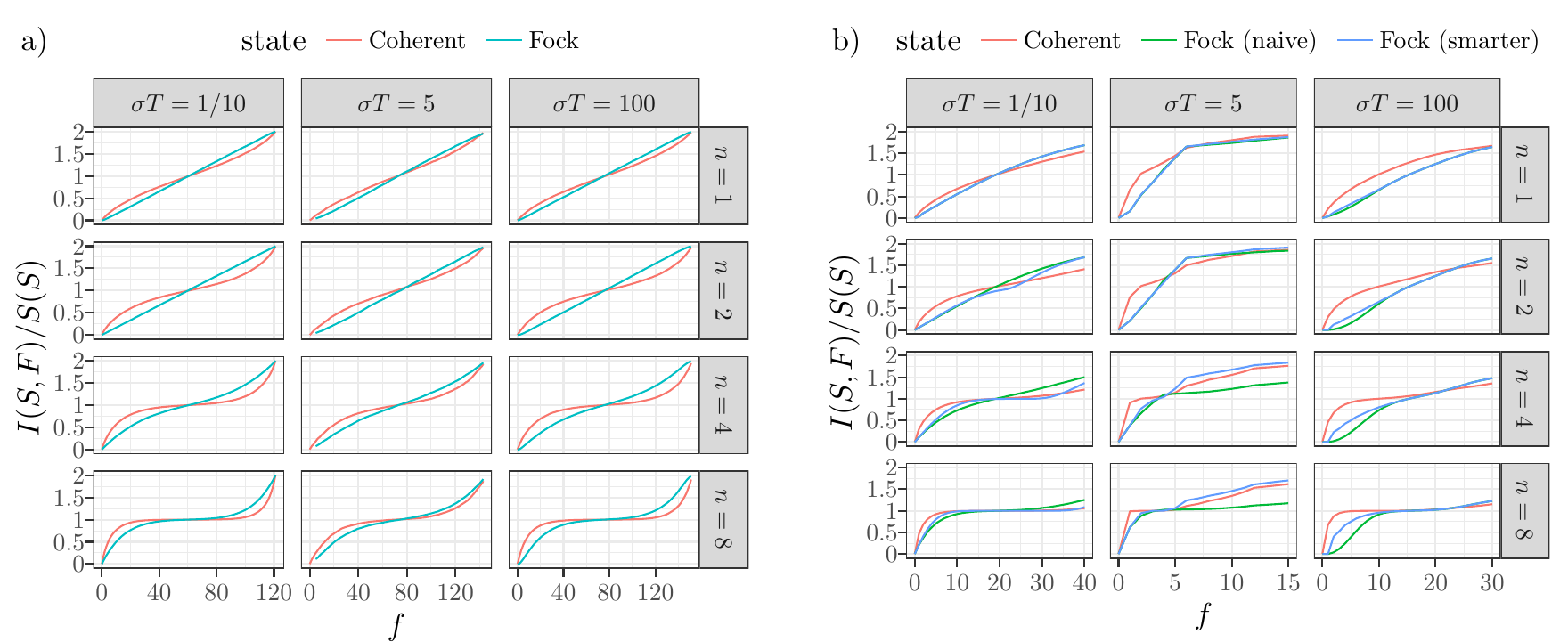}
\caption{(Color online) Quantum mutual information $I(S,F)/S(S)$ for both classical and
quantum probes prepared with a Gaussian wavepacket as a function of the 
size $f$ of the fragment at strong coupling ($\sigma\Delta\tau = 6$).
%
(a) \emph{Random ordering}: 
For $n\leq 2$ no Darwinian plateau is present, independently
of the source whereas it is always present for $n\geq 8$. In the
mesoscopic regime (here $n\simeq 4$), a Darwinian
plateau is present with classical probes but not
with quantum ones.
(b) \emph{Correlation ordering}: 
In the mesoscopic ($n\simeq 4$) time-resolved ($T\sigma=1/10$)
regime, the buildup of the plateau is algorithmic dependent, revealing the
crucial role of correlations.}
\label{fig/mutualinformation}
\end{figure*}

In contrast to the usual models of quantum Darwinism where 
information about the system is uniformly spread among all atomic 
fragments, here it is localized in the time-frequency
plane. Hence, we have to specify how we build our fragments.
To start with, it is always possible to agglomerate atoms by 
randomly choosing them.
This leads to the standard behavior of the (averaged/typical) mutual
information discussed in the usual models of quantum
Darwinism~\cite{Zurek-2009,Riedel-2010,Riedel-2011-1}.
\Cref{fig/mutualinformation}-(a) deals with the strong
coupling limit where the wavepackets $\phi^{(s)}$ for $s=0$, $1$ are
well separated ($\sigma\,\Delta\tau>1$) (for the
weak coupling, see \cref{sec/weak-coupling}). 
The plots depict the ratio $I(S,F)/S(S)$ as a function
of the number of atomic fragments agglomerated for various values of the 
parameter $T\sigma$ and of the average photon number $n$. The red curves
correspond to coherent radiation whereas the blue curves correspond to
Fock states. At low photon number ($n\lesssim 2$), no Darwinian plateau
appears, an expected result since the low photon number prevents
broadcasting of information about $S$ among many fragments. At large
photon number ($n\gtrsim 8$), we see that a Darwinian plateau is present
for both the classical and quantum probes, although less prominent for
the latter. In the intermediate regime ($2\lesssim n\lesssim 4$), 
which we call mesoscopic, we see a Darwinian plateau for classical 
radiation and no plateau for quantum radiation. 

In order to understand this difference, we have to discuss the role of 
correlations between atomic fragments. In this perspective, we 
introduce two algorithms for agglomerating atomic fragments into 
composite fragments of increasing sizes.
The ``naive'' algorithm orders
the atomic fragments $E_{k,l}$ by decreasing order of their correlation with 
the system~\footnote{In case of equality, a random choice is performed.}: 
$I(S,E_1) \geq \dots \geq I(S,E_n)$. 
Composite fragments are then built sequentially by
aggregating atomic fragments accordingly.
Importantly, this construction only requires the reduced density matrices
$\rho_{SE_k}$ and does not use inter-fragment correlations.
By contrast, the ``smarter'' algorithm fully exploits them.
A composite fragment $F_f$ of size $f$ is recursively built by choosing,
at each step, the atomic fragment such that $F_f$ is most correlated
with~$S$.
Given $F_f = \lbrace E_1,\dots,E_f \rbrace$, $F_{f+1}$ is constructed by
aggregating 
\begin{equation}
	E_{f+1} 
=\underset{X}{\mathrm{argmax }}\, I(S,X|E_1,\dots,E_f )
\,
\end{equation}
where $I(S,X|E_1,\dots,E_f )$ denotes the conditional mutual
information between $S$ and $X$ \cite{Book-Wilde}. This algorithm 
requires the whole set of reduced density matrices
$\lbrace \rho_{SE_1\dots E_k} \rbrace$. 

For classical probes, both algorithms lead to the same behavior of 
the mutual information. This comes from the absence of
correlations between the various atomic fragments for a given state of
the qubit. The quantum mutual information between $S$ and any 
fragment $F$ of the environment is then equal to 
(see \cref{sec/details/classical})
\begin{equation}
I(S,F) =h_2(\bigr\lvert D_{\text{tot}} \bigr\rvert)+
f\left(\bigr\lvert D_F \bigr\rvert\right)-
f\left(\left\vert \frac{D_{\text{tot}}}{D_F}
\right\vert\right)\,,
\label{eq/coherentmutualinfo}
\end{equation}
with the binary entropy function
$
h_2(x) = -\frac{1+x}{2} \log\frac{1+x}{2} - \frac{1-x}{2}
	\log\frac{1-x}{2}
$, 
$\lvert D_{\text{tot}}\rvert = |\langle q\phi^{(0)}(\omega)
| q\phi^{(1)}(\omega)\rangle|$ the total decoherence factor and
$|D_F|= \prod_{(k,l) \in F}|\langle q\phi^{(1)}_{kl}
| q\phi^{(0)}_{kl}\rangle|$ the decoherence factor of the fragment
$F$ which is factorized  
into decoherence factors relative to the atomic fragments
within $F$. Since, in the smarter algorithm, 
we look for the atom of signal $X$
maximizing $I(S,E_1,\dots,E_f X)$ or, equivalently,
minimizing 
$D_{S,E_1,\dots,E_f X} = D_{S,E_1,\dots,E_f} D_X$, this
leads to the same atomic fragment than the
naive algorithm which minimizes $D_X$ at each step.

By contrast, in the case of a quantum probe, the difference between the
$I(S,F)/S(S)$ curves obtained with the two algorithms shows the role 
of inter-fragment correlations. In this case, $I(S,F)$ is obtained 
from the reduced density matrix 
$\rho_{SF} =\sum_{k=0}^n p_F(k) \rho_{SF}^{(k)}$
which is an orthogonal mixture with $p_F(k)$ the probability distribution
 to measure $k$ photons in $F$ and $\rho_{SF}^{(k)}$ the conditional 
state upon measuring $k$ photons (see \cref{sec/details/quantum}). 
Consequently, $I(S,F)$ is an average 
$I(S,F) = \mathbb{E}_{p_F(k)}\big( I_{nk}(S,F) \big)$ of
\begin{equation}
I_{n,k}(S,F) =  
	h_2(D_{\text{tot}})
	+ h_2\big( D_{F}(k)\big) 
	- h_2\big(D_{\overline{F}}(n-k)\big)
\,.
\label{eq/Focknk}
\end{equation}
over the photon number probability distribution $p_F(k)$. This quantity
depends on two factors. First the total decoherence 
factor 
$
D_{\text{tot}} = 
	\left(
	\int_0^{+\infty} 
		|\phi(\omega)|^2 \me^{\mi\omega\Delta\tau}
	\,\md\omega
	\right)^n
$
which is the autocorrelation function of the analytic
signal $\phi(t)$ as a function of $\Delta\tau$. Second, the 
``decoherence'' factor relative to the fragment $F$ conditioned to $k$
photons in~$F$, 
\begin{equation}
	D_F(k)
	= \sqrt{1-4(1-|g_F|^{2k})p_F(0|k)p_F(1|k)}\,,
\end{equation}
where
$p_F(s) = \frac{1}{2\pi}\int_{\mathbb{R}^2} W_F(t,\omega)W_{\phi}(t-\tau_s,\omega)
\, \md t\md\omega$ is the probe wavefunction $\phi$
translated at time $\tau_s$ filtered by the window $W_F$ associated with
the fragment $F$ (see \cref{sec/annexe-wigner}). For each $s=0,1$, $p_F(s|k)$ is the
\emph{a priori} conditional probability distribution to have $s\in\lbrace 0,1\rbrace$
knowing $k$ photons were measured. It is
a Poisson distribution with
parameter $p_F(s)$. Finally $g_F = a_F/\sqrt{p_F(0)p_F(1)}$
where
$|a_F|=\left| \frac{1}{2\pi}\int_{\mathbb{R}^2} 
	W_F(t,\omega)W_{\phi}(t-\overline{\tau},\omega) \,\me^{\mi\omega\Delta\tau}
\, \md t\,\md\omega\right|
$ with $\bar{\tau}=(\tau_0+\tau_1)/2$
is the interference pattern between both $s=0,1$ contributions
filtered by the fragment $F$. This is analogous to Mandel's first degree
of coherence for the filtered wavefunction over $F$.

Let us now comment \cref{fig/mutualinformation}-(b). For classical 
probes, both algorithms give the same results but not for quantum 
probes. First of all, the main features 
observed on \cref{fig/mutualinformation}-(a) also appear
for quantum probes: there is no Darwinian plateau at low photon number
($n\lesssim 2$) and a clear plateau appears at large photon number 
($n\gtrsim 8$). 
The physically relevant differences appear in the mesoscopic regime.

In this regime, classical probes have a more visible plateau than
the quantum ones and the difference between both algorithms 
for quantum probes is more visible than at lower or higher photon 
number. This is a consequence of the correlation structure which
we can better witness by discussing the role of the reference frame, 
\textit{i.e.} changing the $\sigma T$ parameter.

In the frequency-resolved case
($\sigma T =100$, right column \cref{fig/mutualinformation}-(b)), a 
Darwinian plateau appears at lower photon number
for classical probes than for quantum probes. This reflects the fact
that coherent states record information in the phase $\omega\tau_s$ 
between the different photon-number states within a single mode,
while quantum radiation does not. 
It is maximal when the phase difference
$\omega\,\Delta\tau$ is $\pi\pmod{2\pi}$ and minimal for
$0\pmod{2\pi}$. This phase can be recovered by homodyne measurements. 
By contrast, Fock states of a fixed $\omega$ mode are eigenstates 
of the number operator $b^\dagger_\omega b_\omega$ which is a building block of 
the dispersive interaction.
Consequently, a Fock state of a frequency-resolved mode cannot decohere 
the system and distinguish between the two classical states. The relative 
phase between two fragments can nevertheless be recovered by considering
interference experiments (see \cref{sec/fock/freq-resolved}). 
This explains why both algorithms develop 
a plateau at larger photon number with the quantum probe than with the 
classical probe as well as the difference at low $f$.

In the intermediate regime $\sigma T=5$, we see a clear difference 
between the two algorithms on \cref{fig/mutualinformation}-(b),
with a very long plateau for the naive one for $n\geq 4$. 
This slower increase of the mutual information arises because the naive 
algorithm aggregates atomic fragments $E_{k,l}$ according to decreasing
order of $I(S,E_{k,l})$. But these are not necessarily the ones that give a fast
rise of the mutual information $I(S,F)$. The smart algorithm is able 
to find an atomic fragment $E_{k,l}$ giving a larger contribution to $I(S,F)$
even if its atomic mutual information $I(S,E_{k,l})$ is lower than for others
$E_{k',l'}$. Additionally, it is also able to recover quantum information 
more efficiently than the naive one 
(see \cref{sec/Holevo} for a discussion of the Holevo information). This
is different from the typical cases of quantum Darwinism where the 
mutual information, the Holevo information and the
quantum discord have symmetry properties~\cite{Zurek-2013}. These
observations question the meaning of the plateau in realistic contexts 
where the information is localized in the environment. 
Indeed, it is not meaningful to interpret the long plateau 
as a redundancy because it may be artificially extended by aggregating
irrelevant atomic fragments. 

In the time-resolved case 
($\sigma T=1/10$, left column of \cref{fig/mutualinformation}-(b)), 
the appearance of the Darwinian plateau at increasing $n$
is expected for both types of probe since, the time delay of the 
outgoing plateau unambiguously determines
the qubit state. Consequently, for increasingly large $n$, more and more
$T\sigma\ll 1$ atoms of signals are able to probe the presence of the 
outgoing pulse. The role of correlations is again
clearly visible: for quantum probes, the ``naive'' algorithm builds a
smaller plateau than with the ``smarter'' algorithm.

Starting from a single atomic fragment, the ``smarter'' algorithm opens 
the time window $F$ around one of the time delays, for instance 
$\tau_0$. Then, the detection of one photon tells us that $s=0$ while
the absence of detection is inconclusive. Consequently, the ``smarter'' 
algorithm builds $F$ to minimize the probability $p_F(0)$ of no 
detection. In this case, $I(S,F)$ is approximately equal to a
classical mutual information associated to a measurement of the 
$\sigma_z$ observable of the qubit and a photon number measurement 
$N_F$ within the time window defined by $F$ (see 
\cref{sec/fock/time-resolved}). Hence, the classical data about~$S$ and
how to access it in~$E$ are fully characterized in this regime. As its 
size $f$ increases, the composite fragment $F$ starts to cover both 
regions around $\tau_0$ and $\tau_1$, leaving an ambiguity on the 
detection time of the photons, a situation observed at lower $f$ with 
the naive ordering algorithm. Then, quantum correlations in the state 
$\rho_{SF}$ are probed and $I(S,F)$ exceeds its
classical bound, ending the Darwinian plateau. 
By exploiting correlations, the ``smarter'' algorithm 
gradually builds a ``path detector'' from photon detection
collecting classical information about $s$
more efficiently than the ``naive'' algorithm.

\emph{Conclusion}
To conclude, we have developed an experimentally relevant model of
a qubit coupled to a transmission line and used time-frequency
analysis to study the resource dependence of quantum Darwinism.
This model, whose predictions may be tested in circuit QED
experiments \cite{Blais-2021-1}, allows to quantitatively analyze the role
of the initial preparation and the choice of reference frame.
Three different regimes of quantum Darwinism
are unraveled: a deeply quantum regime where no classical image can
be reconstructed, a robust Darwinian regime present whatever resources
are used and, in between, a mesoscopic regime where the ability to 
reconstruct a classical picture is resource dependent. 
As emphasized in \cite{Giacomini-2019-1,Le-2020-1,Tuziemski-2020-1}, 
this shows that objectivity is indeed tightly linked to the correlation 
structure between the system and its environment. Interestingly, 
with our realistic model, we can identify the scale where
quantum Darwinism is blurred with a physically relevant parameter 
(photon number). Because the mesoscopic regime occurs at low photon 
numbers, this calls for going beyond asymptotic quantum 
Shannon theory in order to fully understand the 
emergence of objectivity in a quantum world when a single
realization is considered and therefore address the quantum
measurement problem \cite{Book-von-Neumann,Schlosshauer-2005-1}.

\begin{acknowledgments}
We acknowledge support from the European Space Agency (Ariadna Study
1912-01) and the ANR project 
QuSig4QuSense (ANR-21-CE45-0012).
A.F. acknowledges support from ERC grant BLACKJACK (ERC-2019-STG-851866) 
and ANR AI chair BACCARAT (ANR-20-CHIA-0002).
We are thankful to Jean-Michel Raimond, Irénée Frérot and Franck
Laloë for their insightful comments.
\end{acknowledgments}


\bibliography{biblio,biblio/bigbib,biblio/PIC,biblio/livres}

\appendix 

\widetext


\section{Notations and normalizations}
\label{sec/notations}

Starting from mutually commuting bosonic creation and destruction operators $b(\omega)$ for
$\omega>0$ satisfying
\begin{equation}
	[b(\omega),b^\dagger(\omega')]=\delta(\omega-\omega') \,,
\end{equation}
and $\varphi\in L_2(\mathbb{R}^+)$ normalized according to
\begin{equation}
	\int_{\mathbb{R}^+}|\varphi(\omega)|^2\,\md\omega=1
\,,
\end{equation}
we define the creation operator $b^\dagger[\varphi]$ by
\begin{equation}
	b^\dagger[\varphi]=\int_0^{+\infty}\varphi(\omega)\,b^\dagger(\omega)\,\md\omega
	\,,
\end{equation}
which corresponds to creating a single quanta in the single-particle
state $\ket{\varphi}$ such that
$\braket{\omega}{\varphi}=\varphi(\omega)$. The single-particle states
$\ket{\omega}$ for $\omega>0$ are normalized according to
\begin{equation}
	\braket{\omega}{\omega'}=\delta(\omega-\omega')\,.
\end{equation}
Note that with these conventions
\begin{equation}
	\left[b[\varphi_1],b^\dagger[\varphi_2]\right]=\braket{\varphi_1}{\varphi_2}\,,
\end{equation}
where
$\braket{\varphi_1}{\varphi_2}=\int_0^{+\infty}\varphi_1^*(\omega)\,\varphi_2(\omega)\,\md\omega$
denotes the usual hermitian product on $L_2(\mathbb{R}^+)$.
This leads to the equations relating the mode operators $b(\omega)$ to
the $b[\varphi_\alpha]$ associated with an orthonormal basis
$\ket{\varphi_\alpha}$ for $L_2(\mathbb{R}^+)$
\begin{subequations}
	\begin{align}
		b[\varphi_\alpha]&=\int_0^{+\infty}\varphi_\alpha(\omega)^*\,b(\omega)\,\md\omega\\
		b(\omega)&=\sum_\alpha \varphi_\alpha(\omega)\,b[\varphi_\alpha]
	\end{align}
\end{subequations}
which are extensively used in the present paper. The Wigner function
associated with $\ket{\varphi}$ is finally defined as
\begin{equation}
	W_{\varphi}(t,\omega)=\int_{\mathbb{R}}\varphi\left(\omega+\frac{\Omega}{2}\right)
\,,
	\varphi^*\left(\omega-\frac{\Omega}{2}\right)\,\me^{-\mi\Omega
	t}\md\Omega
\end{equation}
which is equivalent to the definition given in the Letter
using the time-domain representation of $\ket{\varphi}$. With this
convention, we have:
\begin{equation}
	\left|\braket{\varphi_1}{\varphi_2}\right|^2=\int
	W_{\varphi_1}(t,\omega)\,W_{\varphi_2}(t,\omega)\,\frac{\md t\md
	\omega}{2\pi}\,.
\end{equation}

\section{Mutual information}
\label{sec/details}

In this appendix, we give a detailed account of the computations of the
mutual information $I(S,F)$ between the system $S$ and a fragment $F$ of
the environment. To have a clearer intuition of the expressions,
we will often write them using a time-frequency
representation. This has the additional advantage of generality since one can write
expressions valid for all wavefunctions $\phi$ and for a general interaction
model controlled by the phase  $\theta_s(\omega)$. The
time-delay model $\theta_s(\omega) = \omega \tau_s$ will be specified
only at the end of the
computations. 

The problem depends on~$5$ parameters:
the type of probe and its intensity, the coupling strength 
and the atomic fragment width and how we build our larger fragments. 
As they all intervene in the analysis of the emergence of objectivity,
let us discuss them more precisely:
\begin{itemize}
\item The input energy $n$ or $\overline{n}=q^2$: both types of radiation have an energy that
can be controlled by the observers, $n$ the number of photons for the Fock
state input and $\overline{n}=q^2$ the intensity of the coherent state input.
We then need to understand what happens at low and high intensities for both
radiations.

\item The time separation $\Delta\tau$: in the 
time-delay model $\theta_s(\omega) = \omega \tau_s$ considered here,
the quantity $\Delta\tau = \tau_1-\tau_0$ measures the
time separation induced by the two different states of the system.
The ratio of this time separation 
$\sigma \Delta\tau$ to the duration of the
probe wavepackets can be viewed as the dimensionless coupling
constant of the problem. The two limiting regimes correspond to well
separated outgoing wavepackets ($\sigma \Delta\tau\gtrsim 1$,
strong coupling)
and strongly overlapping outgoing wavepackets ($\sigma
\Delta\tau\lesssim 1$, weak coupling).

\item The fragment bandwidth $T=1/f$: the time-frequency definition of the
fragments depends on an atom of signal basis parametrized with a width $T$.
There are indeed differences between time-resolved, frequency-resolved and
generic environment decompositions.
\end{itemize}

We first start in \cref{sec/details/classical} by computing the mutual information
when the system is probed by a classical coherent radiation and then 
discuss the quantum Fock probe in
\cref{sec/details/quantum}.
In particular, we give some additional details and toy examples to
better understand what happens in the strong-coupling, time-resolved
case in \cref{sec/fock/time-resolved} and in the frequency-resolved
case in \cref{sec/fock/freq-resolved}

\subsection{Coherent states}
\label{sec/details/classical}

\subsubsection{General setup}

\paragraph*{Entropy of the system}
In the dispersive regime studied here, the outgoing
state of the qubit probed by a coherent state $\ket{\Lambda}$ and 
the environment is an entangled state of the form
\begin{equation}
\ket{\psi_{SE}} = \frac{\ket{0}\ket{\Lambda^{(0)}} + 
	\ket{1}\ket{\Lambda^{(1)}}}{\sqrt{2}}\, ,
\end{equation}
where the coherent states $\ket{\Lambda^{(s)}}$ have a dephased
amplitude conditioned on the state $s$ of the qubit. To compute the
quantum entropy of the system, we need to compute the eigenvalues
of the reduced state $\rho_S$ of the system. In the computational 
basis $(\ket{0},\ket{1})$, it is simply
\begin{equation}
\rho_S= \frac{1}{2}
\begin{pmatrix}
	1& D_{\text{tot}}  \\
	D_{\text{tot}}^*  & 1 
\end{pmatrix}\, ,
\end{equation}
and its eigenvalue are eigenvalues are $(1 \pm |D_{\text{tot}}|)/2$, thereby
leading to
\begin{equation}
	S[\rho_{SE}] = h_2(|D_{\text{tot}}|)
\,,
\end{equation}
in which $h_2$ is the binary entropy function defined by
\begin{equation}
	\label{eq/binary-entropy}
h_2(x) = -\frac{1+x}{2} \log\frac{1+x}{2} - \frac{1-x}{2}
	\log\frac{1-x}{2}.
\end{equation}

\paragraph*{Entropy of the fragments}
Let us now compute the entropy of the fragment $F$ and of $SF$.
We need the eigenvalues of the reduced density matrices.
Using the mode decomposition of a coherent state
\begin{equation}
	\ket{\Lambda^{(s)}} = \bigotimes_{k,l}\ket{\Lambda^{(s)}_{kl}} \, ,
\end{equation}
and the orthogonality relation $\braket{0}{1}=0$, we find
\begin{equation}
\rho_F = \frac{1}{2}\left(
	\ketbra{\Lambda^{(0)}_F}
	+\ketbra{\Lambda^{(1)}_F}
\right)
\,,
\end{equation}
where $\ket{\Lambda^{(s)}_F} = \bigotimes_{(k,l) \in F}\ket{\Lambda^{(s)}_{kl}}$.
Because these states are not orthogonal, the entropy is not equal to \SI{1}{\bit}.
In order to properly compute the entropy of this
state, we decompose it on an orthogonal basis
$\left(\ket{\Lambda^{(0)}_F},\ket{\Lambda^{(1) }_F}^\perp\right)$, obtained from
the Gram-Schmidt orthonormalisation procedure:
\begin{equation}
\ket{\Lambda^{(1)}_F} = D_F \ket{\Lambda^{(0)}_F} + 
	\sqrt{1-|D_F|^2} \ket{\Lambda^{(1) }_F}^\perp
\,,
\end{equation}
in which $D_F = \braket{\Lambda^{(0)}_F}{\Lambda^{(1)}_F}$. The matrix
representing $\rho_F$ in the orthonormal basis $\left(\ket{\Lambda^{(0)}_F},
\ket{\Lambda^{(1) }_F}^\perp \right)$ is then
\begin{equation}
\rho_F  =  \frac{1}{2}
\begin{pmatrix}
1+|D_F|^2 & D_F\sqrt{1-|D_F|^2} \\
D_F^*\sqrt{1-|D_F|^2}  
	& 1-|D_F|^2
\end{pmatrix}\, ,
\end{equation}
which has eigenvalues $(1 \pm |D_F|^2)/2$. This leads to
$S(F) = h_2(|D_F|)$. Finally, the result for $S(SF)$ is
obtained directly by remembering that, the total state for the
system and its environment being pure, $S(SF)=S(\overline{F})$.
Then, $S(\overline{F})$ is computed exactly along the same lines than
$S(F)$ but the decoherence coefficient $D_F$ is replaced by
$D_{\overline{F}}$. Because of the decomposition $\Lambda_E^{(s)}=
\ket{\Lambda_F^{(s)}} \otimes\ket{\Lambda_{\overline{F}}^{(s)}}$, 
it follows that $D_{\overline{F}}=D_{\text{tot}}/D_F$ and this 
leads to $S(\overline{F}) = h_2(|D_{\text{tot}}/D_F|)$.

\subsection{Fock states}
\label{sec/details/quantum}

\subsubsection{General setup}

In the dispersive regime, with an interaction of the form
$-\hbar \sigma_z \otimes \int_0^{+\infty} g(\omega) a^\dagger_\omega
a_\omega \, \md \omega$, an incoming Fock states
$\ket{n_\omega}$ will collect no information
about the qubit since $\ket{n_\omega}$ is an eigenstate of 
$ \int_0^{+\infty} g(\omega) a^\dagger_\omega
a_\omega \, \md \omega$. Indeed,
$\ket{s}\ket{n_\omega} \to
\ket{s}\ket{n_\omega} \me^{\mi (-1)^s g(\omega)\,n_\omega}$
so that $|\langle n_\omega^{(0)} | n_\omega^{(1)} \rangle | =1$.
Therefore,  in order to collect information on the state of the qubit,
the incoming Fock state must involve photons in a specific
wavefunction $\phi(\omega) \in \text{L}^2(\mathbb{R}^+)$,
$\int_0^{+\infty} |\phi(\omega)|^2 \, \md\omega =1$.
We thus consider 
\begin{align}
\ket{n[\phi]} = \frac{\left(a^\dagger[\phi] \right)^n}{\sqrt{n!}}\ket{0}
\, \text{ with }
a^\dagger[\phi] = \int_0^{+\infty} \phi(\omega) a^\dagger(\omega) \,\md\omega
	=\sum_\alpha \phi_\alpha a^\dagger_\alpha
\,
\end{align}
in the mode associated with the normalized wavepacket $\phi$.
In the dispersive regime, the post interaction state is 
the entangled state:
\begin{align}
	\ket{\psi}=\frac{1}{\sqrt{2}}\left(
	\ket{0}\ket{n[\phi^{(0)}]} + \ket{1}\ket{n[\phi^{(1)}]}\right)\,
\end{align}
in which
$\phi^{(s)}(\omega) = \phi(\omega) \me^{\mi\theta_s(\omega)}$
where $\theta_s(\omega)=g(\omega)$.

\paragraph*{Total decoherence factor}
The total decoherence factor $D_{\text{tot}}$ is given by the overlap
between the two relative states of the environment 
$\ket{n[\phi^{(s)}]}$:
\begin{align}
D_{\text{tot}} = \langle n[\phi^{(0)}] | n[\phi^{(1)}] \rangle
= 
	\left(
	\int_0^{+\infty} 
		|\phi(\omega)|^2 \me^{\mi\Delta\theta(\omega)}
	\,\md\omega
	\right)^n
\text{ with }
\Delta\theta(\omega)  = \theta_1 - \theta_0
\,.
\end{align}
It appears as a modulating factor of the coherences in the reduced
density matrix of the system $\rho_S$. In the time-delay model
($\theta_s(\omega) = \omega \tau_s$), we obtain 
\begin{align}
D_{\text{tot}} = 
	\left(
	\int_0^{+\infty} 
		|\phi(\omega)|^2 \me^{\mi\omega\Delta\tau}
	\,\md\omega
	\right)^n
\,,
\label{eq/totaldecofock}
\end{align}
where $\Delta \tau=\tau_1-\tau_0$ denotes the time difference associated with
the two different qubit state. Note that
\begin{align}
D_{\text{tot}} = (G(\Delta\tau))^n
\,,
\end{align}
where
\begin{align}
G(\tau) = \langle \phi^*(t-\tau)\phi(t) \rangle_t
= \int_{\mathbb{R}} \phi^*(t-\tau)\phi(t) \,\md t
\end{align}
is the time-averaged auto-correlation of the time dependent signal
$\phi(t)$ associated
with $\phi(\omega)$.

\paragraph*{Entropy of the system} Let us compute the entropy of the
system. 
It has the same form
as the one obtained for the coherent state input 
$S[S] = f(D_{\text{tot}})$, except that we use 
\cref{eq/totaldecofock} for the total decoherence factor. Once again,
the system spreads one bit of information in its environment when the decoherence
factor is close to zero.

\subsubsection{Fragments and their mutual information}

We now consider the analysis of the output signal with respect
to a decomposition $E = \bigotimes_\alpha E_\alpha$ of the environment
$E$ into a set of atomic elements $E_\alpha$ defined through an atomic
time-frequency decomposition.  We remind that
\begin{align}
a^\dagger[\phi] &= \int_0^{+\infty} \phi(\omega) a^\dagger(\omega) \,\md\omega
	=\sum_\alpha \phi_\alpha a^\dagger_\alpha
\,.
\label{eq/modedecomposition}
\end{align}
Contrary to coherent states which factorize across any mode
decomposition, Fock states in a given mode $\phi$ split according to a
binomial partitioning as we will see now. This generates
correlations between the atomic fragments that have a non-zero overlap
with the mode $\phi$.

A fragment $F$ is again defined as a collection of atomic elements
$E_\alpha$. 
The mode decomposition of \cref{eq/modedecomposition} is then
composed of two parts relative to $F$ and $\overline{F}$ as
\begin{align}
a^\dagger[\phi] = \sum_{\alpha\in F} \phi_\alpha a^\dagger_\alpha
+ \sum_{\overline{\alpha}\in F} \phi_{\overline{\alpha}}a^\dagger_{\overline{\alpha}}
\underset{\text{def}}{=}
a^\dagger_F[\phi] + a^\dagger_{\overline{F}}[\phi]
\,.
\end{align}
From there, we can compute the required reduced density
matrices needed to compute the entropies 
$S(SF)=S(\overline{F})$ as well as $S(S)$ and $S(F)$ to finally obtain
$I(S,F)$.
Let us begin with
the reduced density matrix $\rho_{SF}$:
\begin{equation}
\rho_{SF} = \frac{1}{2}\sum_{k=0}^n \sum_{s,s' \in\lbrace 0,1 \rbrace} \binom{n}{k}
	\left(\sqrt{p_F(s)p_F(s')}\right)^k 
	\langle \phi^{(s')} | \Pi_{\bar{F}} | \phi^{(s)} \rangle^{n-k}
	\ketbra{s}{s'}
\otimes	\ketbra{k,\phi^{(s)}_F}{k,\phi^{(s')}_F}
\,,
\label{eq/reducedmatrixFock}
\end{equation}
where the state
$\ket{k,\varphi}$
is the $k$ photon state in the normalized mode
$\varphi$ and 
\begin{equation}
	p_F(s)=\langle \phi^{(s)}|\Pi_F|\phi^{(s)}\rangle
\end{equation}
as the probability for one photon in the normalized mode $\phi^{(s)}$ to
be detected in $F$ ($\Pi_F$ denoting the orthogonal projector on the
subspace of modes generated by all the atomic modes of the fragments of
$F$). We see that  $\rho_{SF}$ has the form of a
mixture of different photon numbers $N_F=k$ in the fragment $F$:
\begin{equation}
	\rho_{FS}=\mathbb{E}_{p_F(k)}\left[ \rho_{SF}^{(k)}\right]
\end{equation}
where, for clarity, classical averages are denoted by
$\mathbb{E}_{p(x)} \left[O^{(x)}\right] = \sum_x p(x) O^{(x)}$ 
and in which
\begin{equation}
	\label{eq/pfk-definition}
p_F(k) = \binom{n}{k}\,
	\mathbb{E}_{s}\left[ p_F(s)^k
	(1-p_F(s))^{n-k}
	\right]\, ,
\end{equation}
where $\mathbb{E}_s$ denotes the averaging over the balanced
distribution $p(s=0)=p(s=1)=1/2$
and $\rho_{SF}^{(k)}$ is the conditional reduced density matrix when
$k$ photons have been detected. From \cref{eq/reducedmatrixFock,eq/pfk-definition},
$\rho_{SF}^{(k)}$ has the explicit form
\begin{align}
\rho_{SF}^{(k)} = 
\sum_{s,s' \in\lbrace 0,1 \rbrace} 
	\frac{\left(\sqrt{p_F(s)p_F(s')}\right)^k 
	\langle \phi^{(s')} | \Pi_{\bar{F}} | \phi^{(s)} \rangle^{n-k}}{
	\mathbb{E}_s\left[ p_F(s)^k(1-p_F(s))^{n-k}\right]}
	\ketbra{s}{s'}
\otimes	\ketbra{k,\phi^{(s)}_F}{k,\phi^{(s')}_F}\,.
\end{align}
Furthermore, for $k \ne k'$, the conditional
density matrices are orthogonal.

\paragraph*{Entropy of a fragment}
The entropy of a fragment $F$ is obtained by finding the
eigenvalues of the reduced density matrix of $F$ which is obtained from
$\rho_{FS}$ by tracing over $S$:
\begin{align}
\rho_F &= \sum_{k} \binom{n}{k}\,
	\mathbb{E}_{s}\left[p_F(s)^k
	(1-p_F(s))^{n-k}
	\ket{k,\phi^{(s)}_F}\!\!\bra{k,\phi^{(s)}_F}
	\right]
\,.
\end{align}
It is a statistical mixture over the eigenvalues of the photon
number $N_F$ in $F$:
\begin{align}
	\rho_F = \mathbb{E}_{p_F(k)}\left[\rho_F^{(k)}\right]\, ,
\end{align}
with the probabilities $p_F(k)$ given by \cref{eq/pfk-definition} and
\begin{equation}
\rho_F^{(k)} =
	\frac{1}{p_F(k)}\,\binom{n}{k}\,
	\mathbb{E}_{s}\left[
	p_F(s)^k
	(1-p_F(s))^{n-k}
	\ket{k,\phi^{(s)}_F}\!\!\bra{k,\phi^{(s)}_F}
	\right]
\,.
\end{equation}
Since $\rho_F^{(k)} \perp \rho_F^{(k')} = 0$ for $k\neq k'$, the entropy 
can be decomposed into two terms, one coming from the classical
probability distribution $p_F(k)$ and the other from the conditional
density matrix:
\begin{align}
S[\rho_F] = S[p_F(k)] + 
	\sum_{k=0}^n p_F(k) S[\rho_F^{(k)}]
\,.
\end{align}
Because of the orthogonality of the conditioned reduced density
operators $\rho_F^{(k)}$ the Holevo information
of the mixture $\rho_F = \sum_{k=0}^n p_F(k) \rho_F^{(k)}
$ is given by
\begin{equation}
	\chi\left[\left(\rho_F^{(k)},p_F(k)\right)_k\right] = S[\rho_F] -
\sum_{k=0}^n p_F(k) S[\rho_F^{(k)}]
=S[p_F(k)]\,.
\end{equation}
Hence, the quantum information about the system
is entirely contained in the conditional density matrix  $\rho_F^{(k)}$
quantified by its entropy $S[\rho_F^{(k)}]$. This conditioned reduced
density operator can then be written in the
more explicit form
\begin{equation}
\rho_F^{(k)} = \sum_s p_F(s|k)
\ket{k,\phi^{(s)}_F}\!\!\bra{k,\phi^{(s)}_F}\, ,
\end{equation}
with the probabilities $p_F(s|k)$ given by:
\begin{align}
p_F(s|k) &=
\binom{n}{k}p_F(s)^k (1-p_F(s))^{n-k} 
\frac{p(s)}{p_F(k)} \nonumber \\
&=p(s)
	\frac{p_F(s)^k (1-p_F(s))^{n-k} }{\mathbb{E}_{s}
	\left[
	(p_F(s))^k
	(1-p_F(s))^{n-k}
	\right]}
\,.
\end{align}
However
$p(s|k)$ cannot be interpreted as a conditional probability
to obtain $s$ knowing $k$ unless the states $\ket{k,\phi^{(s)}_F}$
are orthogonal, which is not the case in general. The nonorthogonality
of the states $\ket{k,\phi^{(s)}}$ at fixed $k$ may hinder the
estimation of $s$. However, as will be seen in \cref{sec/fock/time-resolved}, the
two states~$\ket{k,\phi^{(s)}_F}$ become orthognal
in the time-resolved case.

After diagonalization of each $\rho_F^{(k)}$ and using their mutual
orthogonality,
the entropy $S[\rho_F]$ is obtained as
\begin{equation}
S[\rho_F] = S[p_F(k)] + 
	\sum_{k=0}^n p_F(k)
	h_2\left(\sqrt{1-4p_F(0|k)p_F(1|k)(1-|g_F|^{2k})}\right)
\,,
\end{equation}
where $h_2$ is the binary entropy function defined by
\cref{eq/binary-entropy}
and $g_F =  \langle \phi^{(0)}|\Pi_F|\phi^{(1)} \rangle/
\sqrt{p_F(0)p_F(1)}$. Note that $g_F$ can be thought of 
as a normalized cross-correlation function in the signal
processing language or as a counterpart of the reduced 
degree of coherence $g^{(1)}$ in optics.

\paragraph*{Mutual information}
Having obtained the entropy for a fragment, we can now
compute the mutual information $I(S,F) = S(S) + S(F) - S(SF)$. Since
the global state is pure $S(SF) = S(\overline{F})$ where $\overline{F}$ is the
complementary fragment of $F$ in the environment.
Note that
$p_F(k) = p_{\overline{F}}(n-k)$, since measuring
$k$ photons out of $n$ in $F$ means that we have $n-k$ in its complementary 
fragment. Finally, we obtain the mutual information as
\begin{equation}
I(S,F) = \mathbb{E}_{p_F(k)}\big( I_{n,k}(S,F) \big)\, ,
\end{equation}
with
\begin{subequations}
\begin{align}
I_{n,k}(S,F) &=  
	h_2(D_{\text{tot}})
	+ h_2\big( D_{F}(k)\big) 
	- h_2\big(D_{\overline{F}}(n-k)\big)\, ,
\label{eq/Ink-definition}
\\
D_{F}(k)
	&= \sqrt{1-4(1-|g_F|^{2k})p_F(0|k)p_F(1|k)}
\,.
\label{eq/Fock-deco-coeff}
\end{align}
\end{subequations}
A few remarks follow from these expressions:
\begin{itemize}

	\item The mutual information $I_{n,k}(S,F)$ is an increasing function of $k$. This is
expected since detecting more photons enables to gather more information
about $S$. To prove this statement, note that
$D_{F}(k) = \sqrt{1-4(1-|g_F|^{2k})p(0|k)p(1|k)}$
is a decreasing function of $k$ because $p(s|k)$ and $1-|g_F|^{2k}$ are increasing
functions of $k$. Consequently, $h_2\big(D_{F}(k)\big)$ and 
$h_2\big( D_{\overline{F}}(n-k) \big)$  are respectively
increasing and decreasing functions of $k$. This implies
		that  $I_{n,k}(S,F)$ is an increasing function of $k$.

\item The mutual information $I(S,F)$ is not the average of the
mutual information obtained from the conditional state 
$\rho^{(k)}_{SF}$ relative to $k$ photons detected in the fragment
$F$. To understand this subtlety, let us recall that $\rho_{SF} = 
\sum_{k=0}^n p_F(k) \rho^{(k)}_{SF}$ and
that $\rho^{(k)}_{SF} \perp \rho^{(k')}_{SF}$ and 
$\rho^{(k)}_F \perp \rho^{(k')}_{F}$ for $k \ne k'$. The crucial point
is that  the
conditional density matrices $\rho^{(k)}_{S}$ are not orthogonal to
		each other. Consequently, $I_{n,k}(S,F)$ defined by
		\cref{eq/Ink-definition} is not equal to
		the mutual information $I[\rho^{(k)}_{SF}]$. The latter quantity
is given by
\begin{align}
I(S,F) &= S[\rho_S] + \sum_{k=0}^n p_F(k) 
	(S[\rho^{(k)}_F] - S[\rho^{(k)}_{SF}]) 
\nonumber \\
	&= \chi\left[\left(p_F(k),\rho^{(k)}_{S}\right)_k\right] + 
\mathbb{E}_{p_F(k)}\big( I[\rho^{(k)}_{SF}] \big)
\,,
\label{eq/mutual-Holevo-decomp}
\end{align}
where, as before, 
$\chi\left[\left(p(x),\rho_{x}\right)_x\right]$ 
denotes the Holevo
information of the ensemble $\lbrace p(x), \rho^{(x)}\rbrace$. 
We recover the fact that the Holevo information is the difference
between the (symmetric) mutual information and what Zurek calls the asymmetric
mutual information
which is 
given by the average of the mutual informations of the conditional
state. Note that this last quantity 
is the conditional mutual information $I(S,F|N_F)$ of $SF$ with respect to
a register storing the number $N_F$ of detected photons in $F$.
Hence, we can write
\begin{align}
I(S,F) = \chi(S, N_F) + I(S,F|N_F)
\,,
\end{align}
		where $\chi(S,N_F)$ is a short notation for the decomposition of
		$\rho_S$ as a statistical mixture of the $\rho_S^{(k)}$ with
		probabilities $p_F(k)$.
\end{itemize}

%
\section{Holevo information}
\label{sec/Holevo}

To properly see that the plateau formed by the mutual information is
composed of classical information, it is useful to compute the Holevo
information. The Holevo information is a bound on the accessible information
which can be used to separate the mutual information into a classical
component and a quantum component also called the quantum discord
\cite{Zurek-2013}. In fact, the Holevo information is in general computed
relative an ensemble $\lbrace (p_x, \rho_x) \rbrace$ where the outcomes
$x$ form a classical random variable $X$. A more physical perspective is
to say that a measurement has been performed on part of the degrees of freedom,
denoted $S$, and an outcome $x$ has been obtained. We then try to evaluate how 
much classical information about the outcome we can obtain from 
measurements on the remaining degrees of freedom $E$.
The Holevo information is a bound on this accessible information. 

In our context, the different ensembles are constructed by choosing 
in which basis we write the state of the system and the relative state
of the field. Up to now, the basis came from the
$\sigma_z$ observable, but can choose any basis, obtained for
instance by rotating the observable on the Bloch sphere. Hence,   
we consider the basis rotated by the spherical angles $(\theta,\phi)$ 
with respect to the $Oz$ axis. The new basis vectors
$\ket{n_{\theta,\phi}}$ satisfy $\sigma_{\theta,\phi} \ket{\pm 
n_{\theta,\phi}} =  \pm \ket{\pm n_{\theta,\phi}}$ where 
$\sigma_{\theta,\phi}$ is the rotated Pauli observable. The Holevo information
has then the form:
\begin{align}
\chi\left[ \left(p(n_{\theta,\phi}), \rho_F(n_{\theta,\phi}) \right)_{n_{\theta,\phi}} \right] 
= S[\rho_F] - p(n_{\theta,\phi})S[\rho_F(n_{\theta,\phi})] 
	- p(-n_{\theta,\phi})S[\rho_F(-n_{\theta,\phi})]
\,,
\end{align}
where the probability distribution $p(\pm n_{\theta,\phi})$ and the
relative states $\rho_F(\pm n_{\theta,\phi})$ will be given below. First, 
we need to write the global state $\ket{\psi}$ of the qubit and the field 
in the new basis. For that, we need:
\begin{subequations}
\begin{align}
\ket{n_{\theta,\phi}} &= \cos\frac{\theta}{2} \, \me^{-\mi \phi/2} \ket{+z} + 
	\sin\frac{\theta}{2} \, \me^{\mi \phi/2} \ket{-z}
\, \\
\ket{-n_{\theta,\phi}} &= -\sin\frac{\theta}{2}  \, \me^{-\mi \phi/2} \ket{+z} + 
	\cos\frac{\theta}{2}  \, \me^{\mi \phi/2} \ket{-z}
\,.
\end{align}
\end{subequations}
The global state can then be written as:
\begin{align}
\ket{\psi} = 
	\sqrt{p(n_{\theta,\phi})} \ket{n_{\theta,\phi}}\ket{\psi(E|n_{\theta,\phi})} +
	\sqrt{p(-n_{\theta,\phi})} \ket{-n_{\theta,\phi}}\ket{\psi(E|-n_{\theta,\phi})}
\,
\end{align}
with $\ket{\psi(E|\pm n_{\theta,\phi})}$ the relative state of the
field given the outcomes $\pm n_{\theta,\phi}$ and $p(\pm n_{\theta,\phi})$
the probability to obtain $\pm n_{\theta,\phi}$. The probability
distribution has the following generic expression:
\begin{align}
p(\pm n_{\theta,\phi}) = \frac{1}{2}
	\left( 1 \pm  
	\text{Re}(\me^{-\mi\phi} D_{\text{tot}}) \sin\theta
	\right)
\,.
\end{align}
The form of the relative states will be given explicitly for each type of probes
from its general definition:
\begin{align}
\rho_F(\pm n_{\theta,\phi}) = \trace_{\overline{F}} 
	\ketbra{\psi(E|\pm n_{\theta,\phi})}
\,.
\end{align}

\subsection{Coherent state}
\label{sec/holev/coherent}

We can now proceed to the explicit computation of the Holevo information.
We start with the coherent state input. The main difference from the
mutual information computation comes from the entropies of the
relative reduced density matrices. We start from the general relative
states:
\begin{align}
\ket{\psi(E|n_{\theta,\phi})} &= 
	\frac{\cos\frac{\theta}{2} \, \me^{-\mi \phi/2} \ket{\Lambda^{(0)}} + 
	\sin\frac{\theta}{2} \, \me^{\mi \phi/2} \ket{\Lambda^{(1)}}}{\sqrt{
	1 +\text{Re}(\me^{-\mi\phi} D_{\text{tot}} \rangle ) \sin\theta}}
\\
\ket{\psi(E|-n_{\theta,\phi})} &= 
	\frac{-\sin\frac{\theta}{2} \, \me^{-\mi \phi/2} \ket{\Lambda^{(0)}} + 
	\cos\frac{\theta}{2} \, \me^{\mi \phi/2} \ket{\Lambda^{(1)}}}{\sqrt{
	1 -\text{Re}(\me^{-\mi\phi} D_{\text{tot}} \rangle ) \sin\theta}}
\,.
\end{align}

\paragraph{Entropy of the fragment} The entropy of the fragment $F$ 
as already been obtained in \cref{sec/details/classical}:
\begin{align}
S[\rho_F] = f(|D_F|)
\,.
\end{align}

\paragraph{Entropy of the relative state} The entropies of the reduced 
relative density matrices are more involved to compute than the one computed for 
the mutual information but proceed along the same lines. To be more
general and be able to apply the result to the Fock state case, let's 
consider the state $\rho = |a|^2\ketbra{\lambda_+}{\lambda_+}
+ |b|^2\ketbra{\lambda_-}{\lambda_-} + 
ab^*\xi\ketbra{\lambda_+}{\lambda_-} + 
a^*b\xi^*\ketbra{\lambda_-}{\lambda_+}$ where $\ket{\lambda_\pm}$ are
two generic state (not necessarily orthogonal) in an Hilbert space. Normalisation
requires that 
$1 = |a|^2 + |b|^2 + 2 \text{Re}(ab^*\xi\langle\lambda_-|\lambda_+\rangle)$.
We want to compute the entropy $S[\rho]$.

As for the mutual information, we again proceed with an orthonormalisation 
procedure and denote the first two orthonomal basis vector
as $\ket{\lambda}$ and $\ket{\lambda^\perp}$. Then we can
rewrite $\rho$ as:
\begin{equation}
\rho  =  \begin{pmatrix}
1-|b|^2 |\braket{\lambda^\perp}{\lambda_-}|^2 &
	|b|^2\braket{\lambda^\perp}{\lambda_-}\braket{\lambda_-}{\lambda}
	a^*b\xi^*\braket{\lambda}{\lambda_-} \\
|b|^2\braket{\lambda^\perp}{\lambda_-}^*\braket{\lambda_-}{\lambda}^*
	ab^*\xi\braket{\lambda}{\lambda_-}^*& 
|b|^2 |\braket{\lambda^\perp}{\lambda_-}|^2 
\end{pmatrix}
\,.
\end{equation}
The probabilities are then obtained as solutions
of the characteristic polynomial:
\begin{align}
p^2 - p + |a|^2|b|^2 (1-|D|^2)(1-|\xi|^2) = 0
\,,
\end{align}
with $D = \braket{\lambda_-}{\lambda_+}$ which are:
\begin{align}
p_\pm = \frac{1}{2}
\left(
	1 \pm
	\sqrt{1-4|a|^2|b|^2 (1-|D|^2)(1-|\xi|^2)}
\right)
\,.
\end{align}
The entropy of this state is then:
 \begin{align}
S[\rho] = f(\sqrt{1-4|a|^2|b|^2 (1-|D|^2)(1-|\xi|^2)})
\,.
\end{align}

We can now apply this formula to compute
$S[\rho(F|\pm n_{\theta, \phi})]$ with 
$4|a|^2|b|^2 = \frac{\sin^2\theta}{4p^2(\pm n_{\theta,\phi})}$,
$|D| = |D_F|$ and $|\xi| = |D_{\bar{F}}|$, so that:
\begin{align}
S[\rho_F(\pm n_{\theta,\phi})] = 
	f\left( 
		\sqrt{1-
		\left(\frac{\sin\theta}{1 \pm \sin\theta \, \text{Re}(\me^{-\mi\phi} D_{\text{tot}})}\right)^2
		(1-|D_F|^2)(1-|D_{\bar{F}}|^2)}
	\right)
\,.
\end{align}

\begin{figure}
	\includegraphics{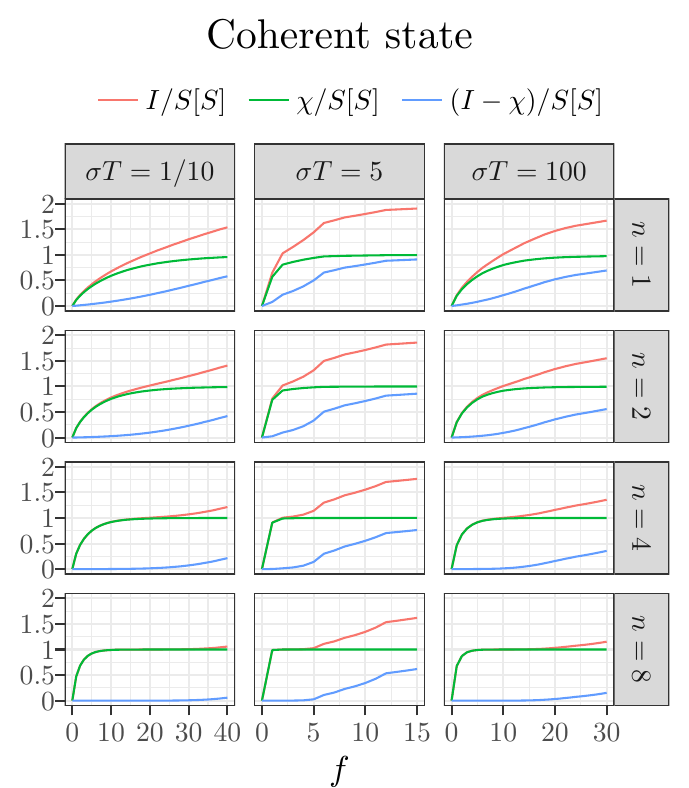}
\caption{Optimal Holevo information for the classical probe in the strong coupling 
regime $\sigma \Delta\tau = 6$.}
\label{fig/holevo-coherent}
\end{figure}

\paragraph{Discussion} \Cref{fig/holevo-coherent} shows the optimal Holevo 
information for the coherent state probe in the different parameter regimes 
explored in the Letter. In all the situations where a plateau is present, the optimal
angles are the physically expected ones $\phi = 0$ and $\theta=0$, \textit{i.e.} the
information about $S$ broadcasted into the environment comes from the
$\sigma_z$ observable. In the microscopic regime, the system does not fully
decohere and the mutual information contains
both classical and quantum information witnessed by the nonzero Holevo information
and discord. However, in the mesoscopic and macroscopic regimes, a plateau
appears and we see that it is indeed essentially composed of classical 
information: the mutual information is approximately equal to the Holevo
information. Furthermore, the departure from the plateau as the size of
the fragment grows is a witness that we start to capture quantum 
correlations which comes from the rise of the discord.

\subsection{Fock state}

We detail here the computations of the Holevo information for
the Fock state input. Once again, we start from the relative states of
the whole environment:
\begin{align}
\ket{\psi(E|n_{\theta,\phi})} &= 
	\frac{\cos\theta/2 \, \me^{-\mi \phi/2} \ket{n[\phi^{(0)}]} + 
		\sin\theta/2 \, \me^{\mi \phi/2} \ket{n[\phi^{(1)}]}}{\sqrt{
	1 +\text{Re}(\me^{-\mi \phi} D_{\text{tot}}) \sin\theta}}
\\
\ket{\psi(E|-n_{\theta,\phi})} &= 
	\frac{-\sin\theta/2 \, \me^{-\mi \phi/2} \ket{n[\phi^{(0)}]} + 
		\cos\theta/2 \, \me^{\mi \phi/2} \ket{n[\phi^{(1)}]}}{\sqrt{
	1 -\text{Re}(\me^{-\mi \phi} D_{\text{tot}} ) \sin\theta}}
\,.
\end{align}
Note that we can obtain the second relative state from the first by the
substitution $\theta + \pi$ into the expression of the first state.
Hence, in what follows, we will focus on the quantum entropy 
coming from $\ket{\psi(E|n_{\theta,\phi})}$.

\paragraph{Entropy of the fragment} The entropy of the
fragment $S[\rho_F]$ was computed in \cref{sec/details/quantum}. It is a sum
of two terms, one given by a classical entropy on the probability distribution
$S[p_F(k)]$ to detect $k$ photons in the fragment $F$ and a second
term which is the average over this distribution of the entropy of
the quantum state of $F$ conditioned on the detection of $k$ photons:
\begin{align}
S[\rho_F] &= S[p_F(k)] + 
	\sum_{k=0}^n p_F(k) S[\rho_F(k)] \\
&= S[p_F(k)] + 
	\sum_{k=0}^n p_F(k)
	f\left(\sqrt{1-4p_F(0|k)p_F(1|k)(1-|g_F|^{2k})}\right)
\,.
\end{align}

\paragraph{Entropy of the relative state} The core of the computation
is essentially the same as the one done for the coherent probe. 
We just need to be careful about the scalar product and the 
decomposition over the photon number. To obtain
the relative state $\rho_F(n_{\theta,\phi})$, we have to expand its definition
$\rho_F(n_{\theta,\phi}) = \trace_{\overline{F}} \ketbra{\psi(E|n_{\theta,\phi})}$,
following the same steps as for the mutual information case. We then
have:
\begin{align}
\rho_F(n_{\theta,\phi}) =  \frac{1}{1 + \sin\theta \Re D_{\text{tot}}}
\Big[ 
	&\sum_{k=0}^n \binom{n}{k} \big( 
	\cos^2\frac{\theta}{2} 
		p^k_F(0) p^{n-k}_{\overline{F}}(0)
		\ketbra{k , \phi^{(0)}} \nonumber \\
	&+ \sin^2\frac{\theta}{2} 
		p^k_F(1) p^{n-k}_{\overline{F}}(1)
		\ketbra{k , \phi^{(1)}} \nonumber \\
	&+ 	\cos\frac{\theta}{2} 	\sin\frac{\theta}{2} \me^{-\mi \phi}
		  \sqrt{p_F(0) p_F(1)}^k
			a_{\overline{F}}^{n-k} \ketbra{k , \phi^{(0)}}{k , \phi^{(1)}}
		+ \text{h.c.}
	\big)
\Big]
\,,
\end{align}
where the amplitude $a_F$ is a short-hand notation for:
\begin{align}
a_F=\langle \phi^{(1)}|\Pi_F|\phi^{(0)}\rangle
\,.
\label{eq/amplitude}
\end{align}
Because of the orthogonality between states with different photon numbers,
we can rewrite this density matrix as a mixture of density matrices 
denoted $\rho_F(n_{\theta,\phi},k)$ conditioned on measuring $k$ photons
(and having prepared the relative state $n_{\theta,\phi})$. We need to 
properly normalize the summands by their traces given by:
\begin{align}
p_F(k|n_{\theta,\phi}) = \frac{\binom{n}{k}}{1 + \sin\theta \Re D_{\text{tot}}}
\left[
	\cos^2\frac{\theta}{2} p^k_F(0) p^{n-k}_{\overline{F}}(0) + 
	\sin^2\frac{\theta}{2} p^k_F(1) p^{n-k}_{\overline{F}}(1) + 
	\sin\theta \Re\left( \me^{-\mi \phi} a_F^k a_{\overline{F}}^{n-k}\right)
\right]
\,.
\end{align}

We can then write $\rho_F(n_{\theta,\phi})$ as a mixture over the distribution
$p_F(k|n_{\theta,\phi})$ of the new relative states $\rho_F(k,n_{\theta,\phi})$ 
conditioned on the new variable $k$ which correspond to measuring
$k$ photons in the fragment $F$. The quantum entropy then expands
into two terms (as in the mutual information case):
\begin{align}
S[\rho_F(n_{\theta,\phi})] = S[p_F(k|n_{\theta,\phi})] + 
	\sum_{k=0}^n p_F(k|n_{\theta,\phi}) S[\rho_F(k,n_{\theta,\phi})]
\,.
\end{align}
To finish the computation, we thus need to evaluate the entropy 
$S[\rho_F(k,n_{\theta,\phi})]$. Given the explicit form of the relative
state $\rho_F(k,n_{\theta,\phi})$, we already obtained the general
form of this entropy in \cref{sec/holev/coherent}.
Using the same notations,  we have
\begin{align*}
4|a|^2 |b|^2 &= \frac{\sin^2\theta}{4 p_F(n_{\theta,\phi},f)} 
	\mathcal{B}(k;n,p_F^{(0)}) \mathcal{B}(k;n,p_F^{(1)}) \\
|D| &= [g_F|^k \\
|\xi| &= |g_{\overline{F}}|^{n-k} 
\,,
\end{align*}
where $\mathcal{B}(k;n,p) = \binom{n}{k} p^k (1-p)^{n-k}$ 
is a short-hand notation for the $k$ binomial distribution among $n$ 
with probability $p$. Hence, this gives
\begin{align}
S[\rho_F(k,n_{\theta,\phi})] =
	h_2\left( 
		\sqrt{1-
		\left(\frac{\sin^2\theta}{4 p_F^2(n_{\theta,\phi},k)} 
			\mathcal{B}(k;n,p_F^{(0)}) \mathcal{B}(k;n,p_F^{(1)})\right)
		(1-|g_F|^{2k})(1-|g_{\overline{F}}|^{2(n-k)})}
	\right)
\,.
\end{align}
We now have all the explicit expressions to compute the
Holevo information over a fragment $F$ when we rotate the state 
of the system by the angle  $(\theta, \phi)$ (or equivalently acquire information
in $F$ about a rotated observable $\sigma_{n_{\theta,\phi}}$ of the qubit):
\begin{align}
\chi\left[ \left(p(n_{\theta,\phi}), \rho_F(n_{\theta,\phi}) \right)_{n_{\theta,\phi}} \right] 
&= S[p_F(k)] - \sum_{n_{\theta,\phi}} p(n_{\theta,\phi}) S[p_F(k|n_{\theta,\phi})] \nonumber \\
&+
	\sum_{k=0}^n p_F(k) S[\rho_F(k)] - 
	\sum_{n_{\theta,\phi}}\sum_{k=0}^n
		p(n_{\theta,\phi}) p_F(k|n_{\theta,\phi}) S[\rho_F(k,n_{\theta,\phi})]
\\
&= \sum_{n_{\theta,\phi}} p(n_{\theta,\phi}) S\big[ p_F(k|n_{\theta,\phi}) || p_F(k) \big] +
	\sum_{k=0}^n p_F(k) \chi\big(\rho_F(k,n_{\theta,\phi}) ; p_F(n_{\theta,\phi}|k) \big)
\,.
\end{align}

\begin{figure}
	\includegraphics{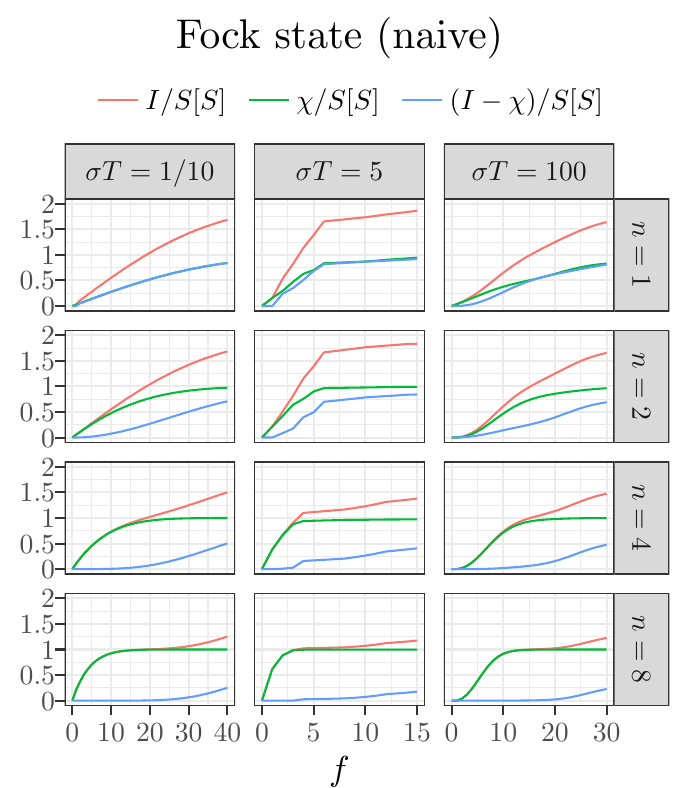}
	\includegraphics{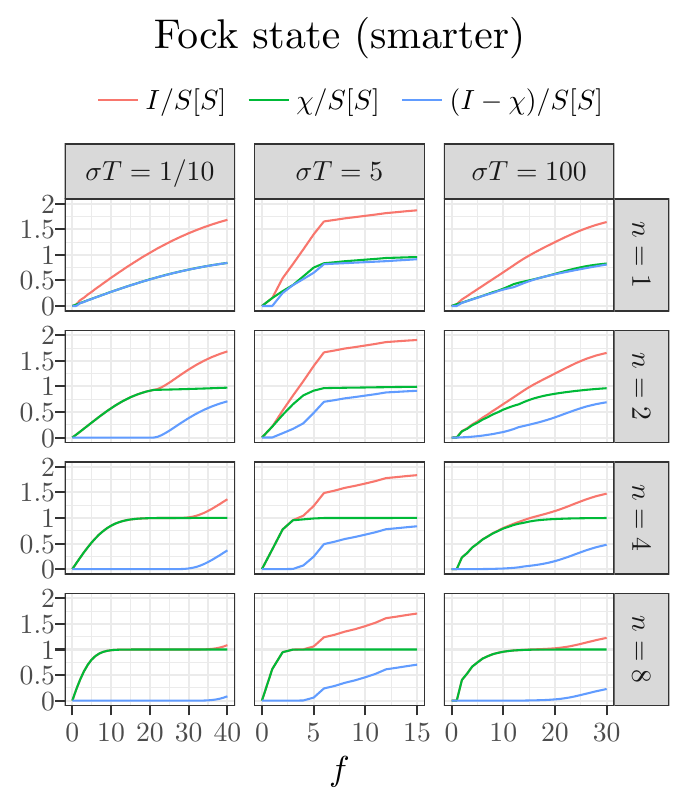}
\caption{Optimal Holevo information for the Fock state probe, in the strong coupling
regime $\sigma\Delta\tau=6$, from the naive algorithm (\emph{left panel})
and the smart algorithm (\emph{right panel}).}
\label{fig/holevo-fock}
\end{figure}

\paragraph{Discussion} \Cref{fig/holevo-fock} shows the optimal
Holevo information for the Fock state probe, for both algorithms, in the parameter
range explored in the Letter. In the plateau regions, the optimal observable is once
again $\sigma_z$. We also have the general feature that the
plateau that appears in the mesoscopic and macroscopic regimes is essentially
classical and that the departure from it comes from the rise of the quantum
discord. The mutual information in the microscopic regime is also a mixture
of both classical and quantum information. Still, we can note that here both 
contributions are essentially equal while, in \cref{fig/holevo-coherent}, the 
classical Holevo information always remains higher than the non-zero discord.

Finally, the comparison between the naive and smart algorithms confirms
what we uncovered in the Letter: the smart algorithm is more able
to recover classical information that the naive algorithm, thanks to
the use of correlations. This is especially clear in the mesoscopic regime
around $n \approx 2$ in the time-resolved region $\sigma T = 1/10$.

\section{Discussion of the various regimes for Fock states}
\label{sec/fock}

In the Letter, we saw that sending a quantum probe required to
take into account subtle correlations in order to access redundantly 
a classical information about the system.
In this appendix, we give some additional details about the different
regimes studied in the Letter for the Fock state input. 
\Cref{sec/fock/freq-resolved} discusses the strong-coupling and
frequency-resolved regime. In particular, we give some elements that
explain that the information about the system is hidden in a phase and
why the ``smart'' algorithm tends to select fragments which are $\pi$-dephased.
\Cref{sec/fock/time-resolved} discusses the time-resolved regime and
we explain in more details why the ``smart'' algorithm tends to build
a fragment around one time-delay first and prove that the mutual
information is indeed classical at first with the photon number as the
optimal measurement to perform in the environment.

\subsection{Strong coupling ($1 \ll \sigma\Delta \tau$) \& Frequency
resolved ($\sigma T \geq 1 $)}
\label{sec/fock/freq-resolved}

\begin{figure}
	\includegraphics[scale=0.8]{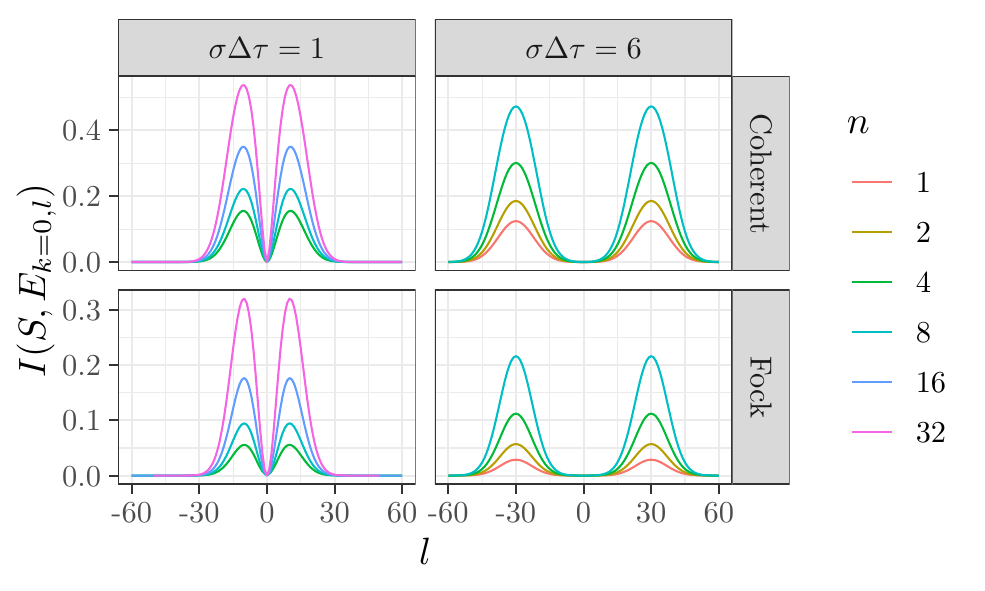}
	\includegraphics[scale=0.8]{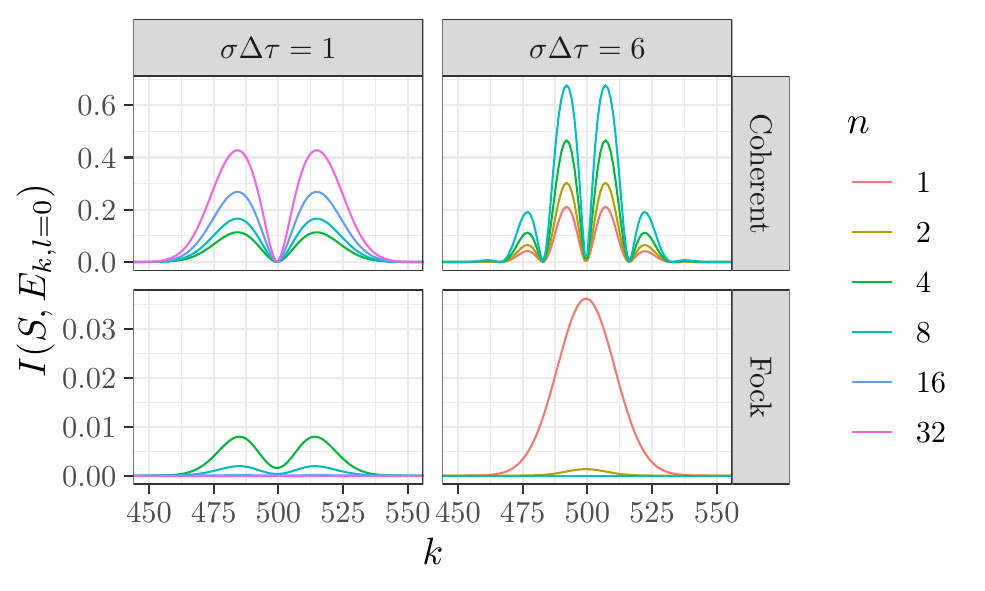}
\caption{Mutual information of atomic fragments in the (relevant part of the)
time-frequency plane in the time-resolved $\sigma T = 1/10$ and 
frequency-resolved cases $\sigma T = 100$. 
\emph{The left panel} shows the mutual information for the atomic
fragment $I_{k=0,l}$ in the time-resolved case. In this case, most of the
information is ``localized at time $l$'' and can be recovered from the first frequency-set
of atoms of signal $k=0$. This situation is very intuitive when we think about
it as a signal with two (Gaussian) components in the time-frequency plane
corresponding to the two possible outcomes $s=0,1$. This is why we have
two spikes in all ranges of parameters. Furthermore, we also see that
the more energy is sent to probe the system, the more information is 
available in each atomic fragments.
\emph{The right panel} shows the mutual information for the atomic
fragment $I_{k,l=0}$ in the frequency-resolved case. This is the complementary
situation where most of the information is ``localized at frequency~$k$'' and can
be recovered from the first time-set of atoms of signal $l=0$. The coherent
case follows the same behavior as before however there are some major
differences in the Fock case.}
\label{fig/mutualinfo-timefreq}
\end{figure}

The right panel of \cref{fig/mutualinfo-timefreq} shows the time-frequency
localization of the mutual information for coherent and Fock states
in the frequency-resolved case $\sigma T = 100 $. In this regime,
almost all the mutual information is located in the fragments situated at
$l=0$, explaining
why only the functions $I_{k,l=0}$ are plotted. The most striking
differences between the two probes appear in the strong coupling regime
$\sigma\Delta\tau =6$.
First, we notice the presence of an oscillation pattern in the time-frequency
plane for the coherent state input. In the frequency-resolved case, the decoherence
factor $D_{k,l=0}$ which controls the value of the mutual information is very
sensitive to the phase difference between $\phi_{k,l=0}^{(0)}$ and 
$\phi_{k,l=0}^{(1)}$. For certain values of $k$, the phase is the same and
therefore we have a decoherence factor $D_{k,l=0}$ equal to one and 
a vanishing mutual information. This phase matching happens when 
$\omega_k\Delta\tau \in 2\pi\mathbb{N}$ which explains the oscillation
pattern. In fact, the time-frequency representations used in \cref{sec/annexe-wigner}
show that this oscillation is quite generic in the frequency-resolved regime
and comes from the cross-terms (interference terms) of the Wigner
function (see for instance \cref{eq/decofactfreqresol}).
Second, we note that for the quantum radiation, contrary to the coherent state
input, the mutual
information in the time-frequency plane decreases in intensity as the
number of photons increases (bottom right graphs of the left panel
of \cref{fig/mutualinfo-timefreq}). The single-photon case is special
here because in this situation, the whole environment of the
qubit is the photon itself. Either it is not detected and nothing can
be inferred about the state of the system or it is detected and there
are no left-out degrees of freedom to decohere the system. All the
information about~$S$ is contained in this photon.
For a larger number of input photons, in the frequency-resolved case, 
most of the information about the system is in the coherences between 
$s = 0$ and $s = 1$. The detection probabilities
are essentially independent of the state of the system. Hence, decoherence
makes it harder to access this information. As we can see from
the general form of the reduced density matrix \cref{eq/reducedmatrixFock},
the factor $\langle \phi^{(s')} | \Pi_{\bar{F}} | \phi^{(s)} \rangle^{n-k}$
decreases as $n$ increases. This qualitatively explains
why, in the frequency-resolved case, the mutual information of the atoms of
signals decreases as $n$ gets higher.

Finally, we can be more precise on how the ``smart'' algorithm
usually works by noticing that the next chosen fragment often has a dephasing
of $\pi$ ($\omega_k \Delta \tau = \pi \pmod{2 \pi}$) compared to the previous one (when the modulus $|\phi(\omega)|^2$
is approximately the same).
To be more quantitative, consider the infinitely frequency-resolved limit with 
atoms of signal indexed by the frequency~$\omega$. In
\cref{sec/annexe-wigner} where the time-frequency Wigner representation
is used, the Wigner function of a plane wave $\me^{\mi\omega_0t}$ is
a Dirac distribution $W_{\omega_0}(t,\omega) = 2\pi\delta(\omega-\omega_0)$. 
Since the dispersive interaction is, by definition, a phase 
shift $\phi^{(s)}(\omega) = \phi(\omega) \,\me^{\mi\theta_s(\omega)}$,
$p_F(s) = p_F$ is independent of $s$. More specifically,
$p_{\omega_0}(s) = \int_{\mathbb{R}} W_{\phi}(t-\tau_s,\omega_0)
\, \md t =|\phi(\omega_0)|^2$. For a large fragment $F$, 
introducing the characteristic distribution $\chi_F(\omega)$ of the 
frequency interval of $F$ leads to write $p_F$ explicitly as:
\begin{align}
p_F(s) = p_F = \int_{\mathbb{R}} \chi_F(\omega) |\phi(\omega)|^2 
\, \md\omega
\,.
\end{align}
It is clear that for any fragment $F$, the probability distribution
$p_F(s)$ does not contain any information about the state of the system.
Consequently, the probability to obtain $k$ photons is then a binomial
distribution $p_F(k) = \binom{n}{k} p_F^k (1-p_F)^{n-k}$
and the conditional mixture probabilities $p_F(s|k) = 1/2$ do not
give any information about~$s$ either. 

Nonetheless, some information about $s$ can be recovered in the overlap
between the two wavepackets. Indeed, for an atom of signal at frequency
$\omega_0$, we have $|a(\omega_0)| = \left| \int_{\mathbb{R}} 
W_{\phi}(t-\bar{\tau},\omega_0) \,\me^{\mi\omega_0\Delta\tau}\, \md t \right| 
=|\phi(\omega_0)|^2$. 
Finally, the decoherence coefficient $D_F(k) = \sqrt{1-4(1-|g_F|^{2k})
p(0|k)p(1|k)}$ simplifies to $|g_F|^k$, like in the coherent state case. 
The mutual information can then be written as:
\begin{align}
	I(S,F) = \mathbb{E}_{p_F(k)}[I_{nk}(S,F)]
\text{ with }
I_{nk}(S,F) = f(D_{\text{tot}}) + f(|g_F|^k) - 
	f(|g_{\bar{F}}|^{n-k})
\,.
\end{align}

\begin{example}
In this frequency-resolved regime, the smart algorithm tends to choose
the next fragment with a dephasing of $\pi$. To qualitatively understand
this, consider the simplest case where $F$ is composed of two modes at
frequency $\omega_1$ and $\omega_2$ respectively. Then, the amplitude 
$a_F$ defined by \cref{eq/amplitude} and the probability $p_F$ are simply:
\begin{align*}
a_F &= |\phi(\omega_1)|^2\me^{-\mi \omega_1 \Delta\tau} + |\phi(\omega_2)|^2\me^{-\mi \omega_2 \Delta\tau} \\
p_F &= |\phi(\omega_1)|^2+ |\phi(\omega_2)|^2
\,.
\end{align*}
We can then compute the normalized factor $g_F$. To do so, it is useful to introduce 
the angle $\cos(\varphi_{12}) = |\phi(\omega_1)| / \sqrt{|\phi(\omega_1)|^2+ |\phi(\omega_2)|^2}$
and the difference $\Delta\omega= \omega_1-\omega_2$.
\begin{align}
|g_F| &= \left| \cos^2 (\varphi_{12}) \, \me^{\mi\frac{\Delta\omega\Delta\tau}{2}} 
	+ \sin^2 (\varphi_{12}) \, \me^{-\mi\frac{\Delta\omega\Delta\tau}{2}} \right|
\nonumber\\
&= \left| \cos\left( \Delta\tau\Delta\omega /2 \right) +
	\mi \cos(2\varphi_{12}) \sin\left( \Delta\tau\Delta\omega /2\right) \right|
\nonumber \\
&=\sqrt{1 - \sin^2(2\varphi_{12}) \sin^2\left( \Delta\tau\Delta\omega /2\right)}
\end{align}
We clearly see that $|g_F|$ is between $0$ and $1$. Besides, if the angle
$\varphi_{12}$ varies slowly, which corresponds to a slowly varying envelope
for the wavefunction, $|g_F|$ is minimal
when $\Delta\tau\Delta\omega = \pi \pmod{2 \pi}$. Hence, given a coupling
$\Delta\tau$ and a wavefunction $\phi$, the lesson from this simple example is
that we should look at modes dephased by a factor of~$\pi$ to recover ``optimally''
the information about $s$.
\end{example}

\subsection{Strong coupling ($1 \ll \sigma \Delta \tau$) \& Time resolved
($\sigma T \leq 1 $)}
\label{sec/fock/time-resolved}

The left panel of \cref{fig/mutualinfo-timefreq} shows the 
mutual information in the time-resolved case  $T \sigma = 1/10$ in
the time-frequency plane. As expected, in the strong-coupling regime
$\sigma\Delta\tau \gg 1$, the information about $s$ is clearly
distinguishable and localized around the times delay $\tau_s/T$ with a width of
$\sigma T$. 

In the strong coupling regime with $\sigma \Delta \tau \gg 1$, it is a good
approximation that the wave functions $\phi^{(0)}$ and $\phi^{(1)}$ are
orthogonal.

The amplitude $a_F$ defined by
\cref{eq/amplitude} thus satisfies $a_F = -a_{\bar{F}}$ so that
$|a_F| = |a_{\bar{F}}|$. In the time-resolved situation 
$\sigma T \leq 1 $, we also have that $|a_F| \approx 0$ in general. 
This can we seen from its time-frequency representation in 
\cref{eq/timefreq-af}. Applied in the extreme time-resolved 
frame, the overlap when $F$ is an atom of signal,
\textit{i.e.} $W_F(t,\omega) = \delta(t-t_0)$, has the form 
$|a(t_0)| = \left| \int_{\mathbb{R}} W_{\phi}(t_0-\bar{\tau},\omega) 
\,\me^{\mi\omega\Delta\tau}\, \md \omega/2\pi \right|$,
or in terms of the wavefunction
$a(t_0)  = \phi(t_0-\tau_0)\phi^*(t_0-\tau_1)$ .
For well-localized wave-packets like Gausssians of width $\sigma$ in
the strong coupling regime, this is almost zero. 
Once again, since the wavepackets $\phi^{(s)}(t)$ are well separated
in time in the strong coupling regime, their overlap will always be close
to zero whatever the window $F$.

Hence, the strongly-coupled and time-resolved regime offers some 
important simplifications on the expression of the mutual information.
With $D_{\text{tot}} \approx 0$ and $a_F \approx 0$, 
\cref{eq/Fock-deco-coeff} simplifies into Shannon's binary entropy
$H_2(p) = -p \log_2(p) -(1-p)\log_2(1-p)$ of the conditional probability 
distributions $(p(s|k))_s$ for $k \ne 0$. The mutual information is obtained
 by summing over the probability
distribution to measure $k$ photons. Most contributions simplify because
of the relation $H_2(p_F(s|k)) = H_2(p_{\overline{F}}(s|n_k))$ except
the extreme cases $k=0$ or $k=n$ (where we have to be careful that
$|a_F|^0 =1$ and consequently $h_2(1) =0$). More explicitly,
\begin{align*}
I(S,F) &= 1 + \sum_{k=0}^n p_F(k) 
\Big( 
	 h_2\big(D_F(k)\big)
	- h_2\big(D_{\overline{F}}(n-k)\big)
\Big) \\
& = 1 + p_F(0) ( 0 - H_2(p_{\overline{F}}(s|n))) + p_F(n) (H_2(p(s|n)) -0)
\,.
\end{align*}
The mutual information in the time-resolved and strongly-coupled regime
is then
\begin{align}
I(S,F) = 1 + p_F(n) H_2(p_F(s|n)) - p_F(0) H_2(p_F(s|0))
\,.
\label{eq/mutual-info-Fock-time-resol}
\end{align}
The formula shows two important things. The first one is that as the 
probability of detecting all the $n$ photons gets higher ($p_F(s) \to 1$
and consequently the no-detection probability goes to zero),
the conditional probability distribution gets balanced, meaning that we
have a stronger uncertainty on which state the system is in and its 
entropy goes to its maximum value of $1$. Hence, the mutual information 
goes over its maximum classical value of $1$. Of course, we have the opposite
behavior if the $p_F(0)$ gets higher  ($p_F(s) \to 0$
and consequently the $n$-photon detection probability goes to zero).
To reach the classical value, we need to be intermediate in the
sense that we need to collect enough photons in the fragment
$F$ to have enough information and leave enough photons out
so that the unmonitored environment still decoheres the system. 

The second important fact is that in this regime, when the window $F$ is
centered around one of the two times $\tau_0$ or $\tau_1$, the conditional
probability distribution $p_F(s|k)$ is the actual conditional
probability distribution to measure $s$ knowing we measured $k$
photons (this will be checked on the state shortly). Hence the mutual
information only depends on the \emph{full counting statistics}
$p_F(k)$; another way to write this in the extremely time-resolved
case is to write $F = [t_1,t'_1] \cup [t_2,t'_2] \cup \cdots$ as a union 
of time intervals and then $p(k ; [t_1,t'_1] \cup [t_2,t'_2] \cup \cdots)$.
To justify all this, consider the conditional state from 
\cref{eq/reducedmatrixFock}: 
\begin{align}
p_F(k)\rho_{SF}^{(k)} &= \frac{1}{2}\sum_{s,s' \in\lbrace 0,1 \rbrace} \binom{n}{k}
	\left(\sqrt{p_F(s)p_F(s')}\right)^k 
		\langle \phi^{(s')} | \Pi_{\bar{F}} | \phi^{(s)} \rangle^{n-k}
	\ketbra{s}{s'}
	\ketbra{k,\phi^{(s)}_F}{k,\phi^{(s')}_F}
\,.
\end{align}
Being time-resolved, $\langle \phi^{(s')} | \Pi_{\bar{F}} | \phi^{(s)} \rangle^{n-k}= 
\delta_{ss'}p_{\overline{F}}(s)$ which
simplifies the expression directly when $k\ne n$ (this is valid even if
$F$ overlaps both spots). When $k=n$,  the simplification
remains thanks to $\sqrt{p_F(s) p_F(s')} = \delta_{ss'}p_F(s)$ which is
only valid when $F$ covers only one of the spot. In short, for all $k$
\begin{subequations}
\begin{align}
p_F(k) \rho_{SF}^{(k)} &= \sum_{s \in\lbrace 0,1 \rbrace} 
	\frac{1}{2}\binom{n}{k}
	p_F(s)^k  (1-p_F(s))^{n-k}
	\ketbra{s}
	\ketbra{k,\phi^{(s)}_F}
\\
&=
\sum_{s \in\lbrace 0,1 \rbrace} 
	p_F(s,k)
	\ketbra{s}
	\ketbra{k,\phi^{(s)}_F}
\,.
\end{align}
\end{subequations}
Since all states in the conditional density matrix are orthogonal,
$p(s,k)$ can be interpreted as the joint probability distribution
to measure the system in the state $s$ and $k$ photons
in the fragment $F$. This consequently justifies the interpretation
of $p_F(s|k) = p_F(s,k)/p_F(k)$. What's more, using the general
decomposition of the mutual information from 
\cref{eq/mutual-Holevo-decomp},
it is straightforward to compute its discord 
$D(S|N_F) = \langle I[\rho^{(k)}_{SF}] \rangle_{p_F(k)}$ relative 
to the measurement the number of photon $N_F$ in the fragment
$F$: 
\begin{align}
D(S|N_F) &= \sum_{k=0}^n p_F(k) 
\big( 
	S[\rho_S^{(k)}] + S[\rho_F^{(k)}] - S[\rho_{SF}^{(k)}]
\big) \nonumber \\
&= \sum_{k=0}^n p_F(k) S[\rho_S^{(k)}]
=0
\,,
\end{align}
with the second line obtained from the fact that $S[\rho_F^{(k)})] =
S[\rho_{SF}^{(k)}] = S[p(s,k]$ and the last equality comes from
the fact that since $F$ is located around either $\tau_0$ or
$\tau_1$, $\rho_S^{(k)}$ is in fact a pure state, either $s=0$
or $s=1$. Hence $I(S,F) = \chi(S, N_F)$ is classical as long as 
$F$ is well time localized and the optimal measurement is the
full counting statistics $N_F$.

\begin{remark}
As a reminder, when we are in the extreme time-resolved frame, the 
probability $p_F(s)$ takes the form:
\begin{align}
p_F(s) = \int_{\mathbb{R}} \chi_F(t) |\phi(t-\tau^{(s)})|^2 
\, \md t
\,.
\end{align}
The probability distribution to measure $k$ photons remains a mixture
of binomial distributions and the conditional probability distributions 
$p_F(s|n)$ and $p_F(s|0)$ that are important for the mutual informaiton
are simply:
\begin{subequations}
\begin{align}
p_F(s|n) &= \frac{p_F^n(s)}{p_F^n(0) + p_F^n(1)} \\
p_F(s|0) &= \frac{(1-p_F(s))^n}{(1-p_F(0))^n + (1-p_F(1))^n}
\,.
\end{align}
\end{subequations}
\end{remark}

\begin{example}
Assume that we have a fragment $F$ that, in the time-frequency plane,
overlaps only a part of the output signal, say the $s=0$ spot, in the strong-coupling
regime. In this situation, we have something close to
\begin{align}
p_F(s) = 
\begin{cases}
    p, & \text{if } s=0 \\
    0, & \text{if } s=1
\end{cases}
\end{align}
This form really embodies the fact that we have well-separated
wavefunction spots in time and that our analysing window probes a partial
part of the $s=0$ spot around time $\tau_0$ and not at all the $s=1$ spot
at time $\tau_1$. Then the probability $p_F(k)$ which is a mixture of 
binomial distributions given in \cref{eq/pfk-definition} is:
\begin{align}
p_F(k) = 
\begin{cases}
    \frac{1}{2} \binom{n}{k}p^k(1-p)^{n-k}, & \text{if } k \ne 0 \\
    \frac{1}{2}((1-p)^n + 1), & \text{if } k=0
\end{cases}
\end{align}
The distinction between the case of detection of some photons with no
detection is important. Indeed, when no photons are detected, we do not
know if it is because of the quantum randomness of the measurement outcome
or if it is because there was no signal in the first place.
The form of the conditional probability distribution $p(s|k)$ shows this
in a transparent way:
\begin{align}
p_F(s|k) = 
\begin{cases}
\text{if } k \ne 0 &
   	\begin{cases} 
	1, & \text{if } s=0 \\
	0, & \text{if } s=1
	\end{cases}, 
\\
\text{if } k = 0 &
   	\begin{cases} 
	\frac{(1-p)^n}{(1-p)^n + 1}, & \text{if } s=0 \\
	\frac{1}{(1-p)^n + 1}, & \text{if } s=1
	\end{cases}
\end{cases}
\end{align}
Hence, if some photons are detected, even one, in the time-resolved
window around $\tau_0$, we are then sure that the system was in the
state $s=0$. It is a consequence of the strong-coupling regime and the
fact that the measurement window does not overlap the time interval 
around $\tau_1$. A detection can perfectly distinguish between the
two states. However, when no photons are detected, there is an 
ambiguity coming, as we said, from the fact that maybe there was no
signal in this time interval in the first place (system in state $s=1$) or
or maybe it is a consequence of quantum randomness. Naturally,
we see that when $p$ is close to $1$ (window $F$ of the size of the
spot), the no-detection event in this time interval is really improbable
which implies that we can be almost sure, even sure when $p=1$, that
the system is in the state $s=1$: we have a good measurement device.
If, however, $p$ is close to $0$ (either because the window is too small
so we have a small amount of energy or because we probe
at the wrong time), the no-detection event gives us nothing more
then a pure random fifty-fifty guessing of the state of the system.

In this case, because any detection of photon perfectly distinguishes the
state of the system so that $p(s|k\ne 0)$ is binary, the mutual information 
\eqref{eq/mutual-info-Fock-time-resol} simplifies into:
\begin{align}
I(S,F) = 1 - p_F(0) H_2(p_F(s|0))
\,.
\end{align}
This expression makes it clear that no-detection events (no-signal or dark counts) 
hinder the amount of classical information (either $s=0$ or $s=1$) we can
extract from output signal. As the probability $p_F(0)$ is higher, up to $1$
($p \to 0$), the conditional probability $p(s|0)$ gets more balanced and its entropy 
gets higher, up to $1$. Hence, the mutual information goes down to zero which means
that our filtered signal is useless to get information about the state of the system.
\end{example}

\begin{example}
Assume that the window $F$ is large enough so that it overlaps both output
spots equal. Concretely we can imagine a measurement device that probes the
signal equally around $\tau_1$ and $\tau_0$. In this situation, we have simply
\begin{align}
p_F(s) = p_F
\,.
\end{align}
Then it is straightforward to obtain that the probability $p_F(k)$ to 
detect $k$ photons is a binomial:
\begin{align}
p_F(k) = \binom{n}{k}p_F^k(1-p_F)^{n-k}
\,,
\end{align}
and that the conditional probability distribution $p_F(s|k) = 1/2$ does
not depend on $s$. Its entropy is of course maximal 
$H_2(p(s[k)) = 1$. In this case the mutual information simplifies
drastically into:
\begin{align}
I(S,F) = 2p_F
\,.
\end{align}
This example is by construction ambiguous in the possibility to 
distinguish the two states $s=0,1$. Indeed, the time-resolved measurement
device works on both time intervals where the signal is but we do not
know for sure from which spots the detected photons came from. Hence
the fact that the conditional probability $p_F(s|k)$ is flat. Note finally that
if the window $F$ is sufficiently large to cover the whole signal, \textit{i.e.} we measure
the whole environment, we naturally obtain the maximal value of $2$ of the
mutual information.
\end{example}

\section{Weak coupling ($1 \gg \sigma \Delta\tau$)}
\label{sec/weak-coupling}

\begin{figure}
	\includegraphics{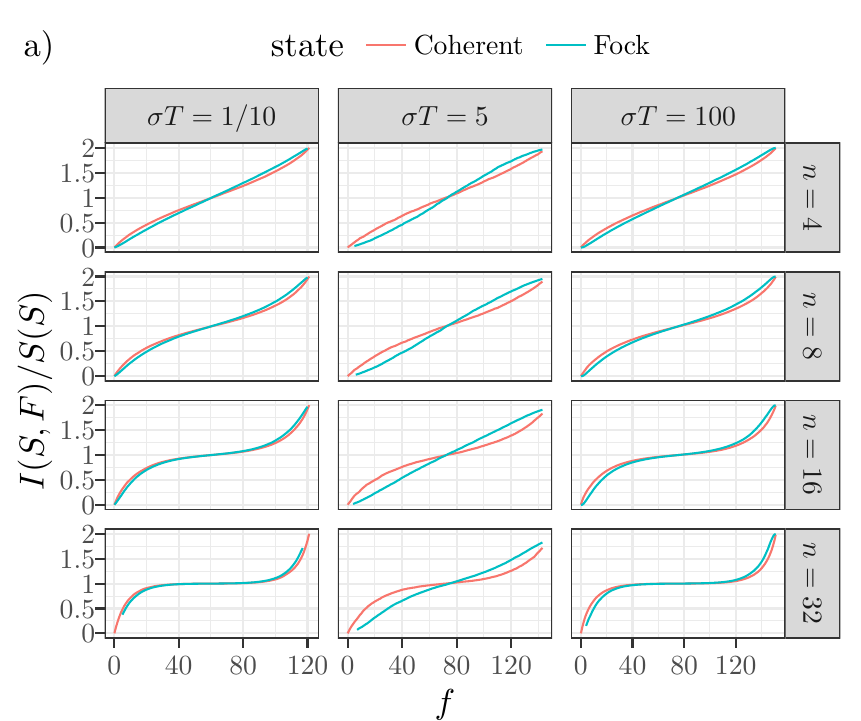}
	\includegraphics{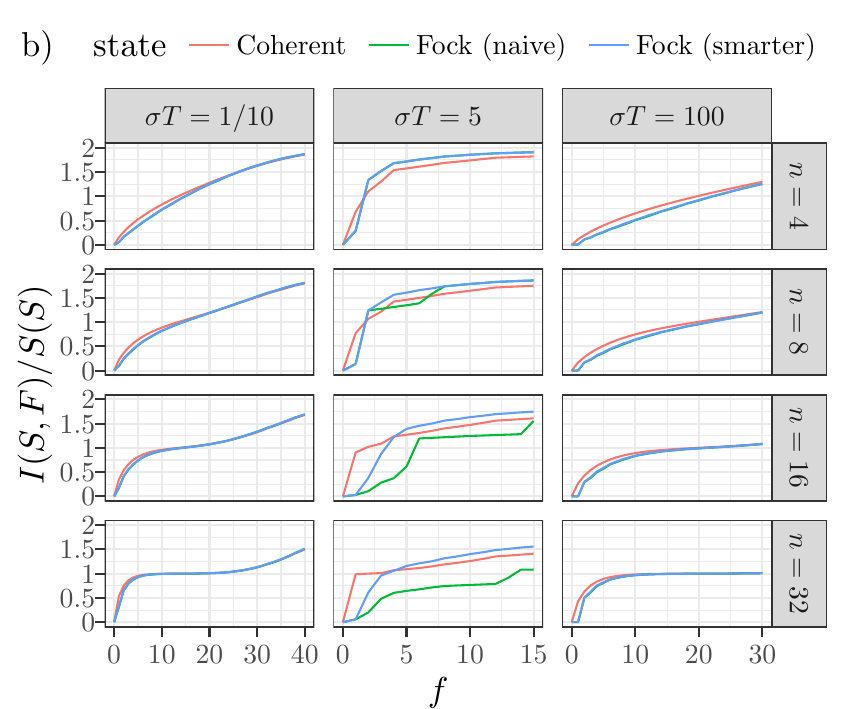}
\caption{Averaged quantum mutual information for different input states (classical and
quantum probe) for different values of the intensity $n$ and time-frequency
width $T$, in the weak-coupling regime $\sigma\Delta\tau = 1$.
\emph{Left panel}: The fragments are chosen randomly among the atomic fragments. 
The usual behavior of quantum Darwinism is present, especially the symmetry around half
the size of the environment. The presence of the plateau depends on the type of 
probe in a mesoscopic regime which is more narrow in time-frequency ($\sigma T \approx 1$)
compared to the strong-coupling case and is at higher intensity $n$.
\emph{Right panel}: The fragments are chosen according to the two
different algorithms, the naive and correlation-based ones. Non trivial features
appears in the same mesoscopic regime revealed by the random algorithm.
We again see that, when the system is probed with a quantum source, the classical 
plateau is best recovered when correlations between the atomic fragments are
taken into account. Note however that the plateau appears
at a much higher intensity $n$. This can be understood from the fact that the overlap
between the two signals $\phi^{s}$ is close to $1$ as a consequence of the weak
coupling assumption. But decoherence is needed for objectivity to appears, with 
$D_{\text{tot}} \approx 0$. This is achieved by sending more photons to probe
the system.
}
\label{fig/randommutualinformation}
\end{figure}

The Letter focused on analyzing the model in the
strong-coupling regime. To be complete, we present here
the same results in the weak-coupling regime.
\Cref{fig/randommutualinformation} shows 
the averaged mutual information for different photon numbers,
different filtering width for both coherent and Fock states inputs
in the weak coupling regime and compares it to the correlation-based
algorithms. 

The left panel of \cref{fig/randommutualinformation} 
shows the mutual information when the fragments
are built from a random choice of atoms of signal. It already shows
some qualitative differences compared to the strong coupling regime.
First, that in both the frequency-resolved and time-resolved cases,
the type of probe sent does not affect the mutual information as a
function of the size  of the fragment. This is however not the case in the
intermediate region $\sigma T \approx 1$. Moreover, we see again
that, as a function of the input intensity $n$, we can distinguish
three cases, a ``quantum or microscopic'' regime with a low number of 
photons and no classical plateau, a ``macroscopic'' regime with a 
high number of photons and a large classical plateau  and finally
a ``mesoscopic'' regime where the type of probe plays an important
role for the existence or not of a classical plateau. Note the major
difference in the number of photons that has to be sent ($n \approx$ 16 or 32).
This can be understood from the generic form of the decoherence
factor 
$D_{\text{tot}} = \langle n[\phi^{(0)}] | n[\phi^{(1)}] \rangle
= 
	\left(
	\int_0^{+\infty} 
		|\phi(\omega)|^2 \me^{\mi\Delta\theta(\omega)}
	\,\md\omega
	\right)^n
\text{ with }
\Delta\theta(\omega)  = \omega\Delta\tau$. For a Gaussian
wavepacket, the total decoherence factor is also Gaussian with
an argument scaling as $n \Delta\tau^2$. If we want to compare
the physics of the strong versus the weak coupling regimes, we should
be at roughly the same decoherence factor, hence 
$n_w \Delta\tau_w^2 = n_s \Delta\tau_s^2$ where $n_{w/s}$ and
$\Delta\tau_{w/s}$ are the photon number and the effective 
coupling constant for the weak and strong coupling situation respectively.
We then have an inverse quadratic scaling for the photon number
$n \propto 1/\Delta\tau^2$ which tends to diverge at low coupling.
All in all, the main lesson of the random choice situation is that
something non trivial happens again in a mesoscopic regime which
is to be found in a more constrained time-frequency region and at
higher input intensity. 

Given this information, we can now explore the problem anew
by using the correlation-based algorithms. The right panel
of \cref{fig/randommutualinformation} shows again the mutual
information in the weak coupling regime. The non-trivial regime
is confirmed to be in the mesoscopic region identified above. 
In fact, as in the strong coupling case, we see that correlations
play a crucial role to recover a classical information. In the 
regime where a plateau is present with a coherent state probe, 
a plateau can only be recovered with the Fock probe if the fragments
are built from the ``smart'' algorithm. Meanwhile, the ``naive''
algorithm is unable to recover enough information (see for
instance the case $\sigma T = 5$ and $n=32$, where the ``naive''
algorithm slowly reaches the classical bound $I(S,F)/S(F) =1$).

To sum up, the weak coupling situation shows again the
important role of correlations in the reconstruction process.
However, compared to the strong coupling situation, the region
where this is relevant becomes narrower in the parameter
space and demands to input more energy in the system. This is
physically reasonable since that in order to have good decoherence, 
enough entanglement must exist between the system, the observer 
and the rest of the environment. Consequently, these systems must
interact enough for these correlations to appear. By definition of the
weak coupling regime, the interaction is quite low there which explains
why it is harder to get a good decoherence and to recover
the classical data.

\section{Time-frequency analysis}
\label{sec/annexe-wigner}

In this appendix, we discuss the time-frequency
representation of the different decoherence factors and of the
probabilities that appear in the expression of the 
quantum information quantities used in the Letter. The main 
advantage of this Wigner-like representation is that it gives a clear graphical
way to to understand the qualitative features of the decoherence factors.

\subsection{Coherent states}

\paragraph{General forms}

All the decoherence factors that appear in the mutual and Holevo informations
are the squared norm of a scalar product between our atom of signal and 
the output wavefunction. Generally, such a scalar product between functions
is nicely represented in terms of the Wigner representation of these functions
using Moyal's identity
\begin{align}
|\langle x | y \rangle|^2 = 
	\int_{\mathbb{R}^2} 
		W_x(t,\omega) W_y(t,\omega)
	\, \md t \frac{\md\omega}{2\pi}
\,.
\end{align}
With the Wigner representation, the exponents of the total decoherence 
factors, $|D_F|= \prod_{(k,l) \in F}|\langle q\phi^{(1)}_{kl}
| q\phi^{(0)}_{kl}\rangle|$, are proportional to 
overlap between the Wigner representations of the two delayed 
wavepackets and the Wigner function of the fragment $F$. For an
atom of signal $(k,l)$, we have
\begin{align}
\log(|D_{(k,l)}|) &= -\frac{1}{2}
	\int_{\mathbb{R}^2} 
		W_{k,l}(t,\omega)W_{\phi^{(0)}-\phi^{(1)}}(t,\omega)
	\, \md t \frac{\md\omega}{2\pi}
\\
&=-\frac{1}{2} 
	\int_{\mathbb{R}^2} 
		W_{k,l}(t,\omega)
		\big(W_{\phi^{(0)}}(t,\omega) 
		+ W_{\phi^{(1)}}(t,\omega) 
		-2\text{Re}(W_{\phi^{(0)},\phi^{(1)}})(t,\omega)
		\big)
	\, \md t \frac{\md\omega}{2\pi} \,.
\end{align}
This expression is well suited for an interpretation in the time-frequency
space. An important subtlety, however, comes from the cross-Wigner function
term~$W_{\phi^{(0)},\phi^{(1)}}(t,\omega)$, which encodes the interference
between the two signals.

In our model, the signals~$\phi^{(s)}$ are obtained by a time
translation of the original signal by a delay~$\tau_s$. In
this case, the cross Wigner function is the Wigner function
of the original signal $W_\phi$ delayed by the average time~$\overline{\tau} =
(\tau_1 + \tau_2)/2$ and
modulated by a frequency-varying phase proportional $\Delta\tau$. 
For our Gaussian signals, the time-frequency representation of the output
signal is made of three parts. Two Gaussian spots correspond to the
signals~$\phi^{(s)}$. The third part is another Gaussian spot, modulated in
frequency, corresponding to the interference term. It encodes the phase between
the two wavepackets.
The decoherence factor is then written as:
\begin{align}
\log(|D_{(k,l)}|) &=-\frac{1}{2} 
	\int_{\mathbb{R}^2} 
		W_{k,l}(t,\omega)
		\big(W_\phi(t-\tau_0,\omega) 
		+ W_\phi(t-\tau_1,\omega) 
		-2 \cos(\omega\Delta\tau)  W_\phi(t-\overline{\tau},\omega)
		\big)
	\, \md t \frac{\md\omega}{2\pi}
\,.
\label{eq/timefreqrepdecoherencecohernet}
\end{align}
Furthermore, we write the decoherence
factor $D_F$ of a fragment $F$ composed of many atoms of
signals. Indeed, thanks to the tensor product
structure of coherent states, $D_F$ is the product of the
decoherence factors of its elements. In terms of Wigner
function, this means that the Wigner function of the fragment
$W_F$ is the sum of the Wigner function of its elements:
\begin{align}
W_F(t,\omega) = \sum_{(k,l)\in F} W_{k,l}(t,\omega)
\,.
\end{align}
Thus:
\begin{align}
\log(|D_F|) &=-\frac{1}{2} 
	\int_{\mathbb{R}^2} 
		W_F(t,\omega)
		\big(W_\phi(t-\tau_0,\omega) 
		+ W_\phi(t-\tau_1,\omega) 
		-2 \cos(\omega\Delta\tau)  W_\phi(t-\overline{\tau},\omega)
		\big)
	\, \md t \frac{\md\omega}{2\pi}
\,.
\end{align}
As an example, we can rewrite the total decoherence factor, by considering the
fragment~$F$ to be composed of all the atoms of signal,
\begin{align}
\log(|D_{\text{tot}}|) &=-\frac{1}{2} 
	\int_{\mathbb{R}^2} 
		(W_\phi(t-\tau_0,\omega) 
		+ W_\phi(t-\tau_1,\omega) 
		-2 \cos(\omega\Delta\tau) W_\phi(t-\overline{\tau},\omega)
	\, \md t \frac{\md\omega}{2\pi}
\nonumber\\
&=-\int_{\mathbb{R}^2} 
		W_\phi(t,\omega) 
		(1-\cos(\omega\Delta\tau))
	\, \md t \frac{\md\omega}{2\pi}
\underset{\int_{\mathbb{R}}W_\phi(t,\omega) \, \md t = |\phi(\omega)|^2}{=}
-\int_{\mathbb{R}} 
	|\phi(\omega)|^2
	(1-\cos(\omega\Delta\tau))
	\, \frac{\md\omega}{2\pi}
\,.
\end{align}

\paragraph{Frequency-resolved case}
This time-frequency representation allows to clearly
understand the origin of the oscillations of the mutual
information in the frequency-resolved case seen in
\cref{fig/mutualinfo-timefreq}. Indeed, consider a 
frequency-resolved atom of signal such that its Wigner
representation is given by 
$W_{k,l}(t,\omega) \approx 2\pi \delta(\omega-\omega_k)$.
A direct application of \cref{eq/timefreqrepdecoherencecohernet}
gives
\begin{subequations}
\label{eq/decofactfreqresol}
\begin{align}
\log(|D_{(k,l)}|) & \approx -\frac{1}{2} 
	\int_{\mathbb{R}} 
		\big(W_\phi(t-\tau_0,\omega_k) 
		+ W_\phi(t-\tau_1,\omega_k) 
		-2 \cos(\omega_k\Delta\tau)  W_\phi(t-\overline{\tau},\omega_k)
		\big)
	\, \md t
\\
&\approx
	-\int_{\mathbb{R}}
		W_\phi(t,\omega_k) 
		\big(1- \cos(\omega_k\Delta\tau) \big)
		\big)
	\, \md t
=-2\int_{\mathbb{R}}
		W_\phi(t,\omega_k) 
		\sin^2(\omega_k\Delta\tau/2)
	\, \md t
\\
&\approx
-2\sin^2(\omega_k\Delta\tau/2) |\phi(\omega_k)|^2
\,.
\end{align}
\end{subequations}
The oscillation of the decoherence factor and consequently
of the mutual information clearly appear as the consequence
of interference pattern of the two delayed signal (cross-Wigner
term). We recover that when 
$\omega_k \Delta\tau/2 \in \pi\mathbb{N}$ or
$kf\Delta\tau \in \mathbb{N}$, the decoherence factor
$|D_{k,l}|$ is approximately equal to one, giving a zero
mutual information.

\paragraph{Time-resolved case}
The opposite case of a perfectly time-resolved analysis 
can be understood in simple terms with the time-frequency
representation. Indeed, consider a time-resolved atom of signal
such that its Wigner representation is given by 
$W_{k,l}(t,\omega) \approx \delta(t-t_l)$.
\Cref{eq/timefreqrepdecoherencecohernet} then gives:
\begin{align}
\log(|D_{(k,l)}|) 
&\approx
-\frac{1}{2} 
	\int_{\mathbb{R}} 
		\big(W_\phi(t_l-\tau_0,\omega) 
		+ W_\phi(t_l-\tau_1,\omega) 
		-2 \cos(\omega\Delta\tau)  W_\phi(t_l-\overline{\tau},\omega)
		\big)
	\, \frac{\md\omega}{2\pi}
\\
&
\approx 
-\frac{1}{2}
\big(
	|\phi(t_l-\tau_0)|^2 + |\phi(t_l-\tau_1)|^2
	-\phi^*(t_l-\tau_0)\phi(t_l-\tau_1) - \text{h.c.}
\big)
\\
&
\approx 
-\frac{1}{2} \big| \phi(t_l-\tau_0) - \phi(t_l-\tau_1) \big|^2
\,.
\end{align}
In the time-resolved case, we recover a very common
form of the decoherence factor. Indeed, it has an exponential
form and the exponent depends on ``the distance'' between 
the two wavepackets.

\subsection{Fock states}

The same reasoning applies for the Fock state input.
Two quantities appear in the reduced density matrix and were denoted
$p_F(s)$ and $a_F = \langle \phi^{(1)}|\Pi_F|\phi^{(0)}\rangle$. From 
\cref{eq/reducedmatrixFock}, we see that the diagonal elements are
controlled by~$p_F(s)$ only. Since this quantity is a scalar product, we can
express it in terms of Wigner representations,
\begin{align}
p_F(s) = \sum_{\alpha \in F} |\phi_F^{(s)}|^2
= &\int_{\mathbb{R}^2} W_F(t,\omega)W_{\phi^{(s)}}(t,\omega)
\, \md t \frac{\md\omega}{2\pi} \\
\underset{\theta(\omega)=\omega\tau_s}{=}
&\int_{\mathbb{R}^2} W_F(t,\omega)W_{\phi}(t-\tau_s,\omega)
\, \md t \frac{\md\omega}{2\pi}
\,.
\label{eq/timefreq-pf}
\end{align}
We see that the probability $p_F(s)$ is given as an average
value over the whole time-frequency plane of the Wigner function of 
the signal $\phi^{(s)}$.

Similarly, the amplitude $a_F = \langle \phi^{(1)}|\Pi_F|\phi^{(0)}\rangle$
is a scalar product and can be written with the Wigner representations as
\begin{align}
|a_F|=\left|\trace\left(\sum_{\alpha\in F} \ketbra{\alpha} 
	\ketbra{\phi^{(0)}}{\phi^{(1)}}\right)\right|
=&\left| \int_{\mathbb{R}^2} 
	W_F(t,\omega)W_{\phi^{(1)},\phi^{(0)}}(t,\omega)
\, \md t \frac{\md\omega}{2\pi}\right| \\
\underset{\theta(\omega)=\omega\tau_s}{=}
&\left| \int_{\mathbb{R}^2} 
	W_F(t,\omega)W_{\phi}(t-\overline{\tau},\omega) \,\me^{\mi\omega\Delta\tau}
\, \md t \frac{\md\omega}{2\pi}\right| \,.
\label{eq/timefreq-af}
\end{align}
This amplitude is given by the average over
the time-frequency plane of the crossed Wigner function of the 
two signals $\phi^{(0)}$ and $\phi^{(1)}$ weighted by the Wigner function
of the $F$. Note again that given our time-delayed model, the crossed 
Wigner function simplifies to the Wigner function of the input signal 
$\phi$ itself evaluated at the mid-point $\overline{\tau} = (\tau_0
+\tau_1)/2$ times a modulation in $\Delta \tau = \tau_1-\tau_0$.

Hence, the diagonal part of the density matrix is controlled 
by the two signal spots while the coherences are controlled 
by the interference pattern. Note that this separation into two
quantities, one from the interference term from~$a_F$ and the other
from the intensity of the signal from~$p_F$, is different from
the coherent state case, where both the signals and the interference
terms contribute to the decoherence factors 
\cref{eq/timefreqrepdecoherencecohernet}.

\end{document}